\documentclass{article}

\PassOptionsToPackage{numbers}{natbib}
\usepackage[final]{neurips_2021}

\usepackage[utf8]{inputenc} % allow utf-8 input
\usepackage[T1]{fontenc}    % use 8-bit T1 fonts
\usepackage{hyperref}       % hyperlinks
\usepackage{url}            % simple URL typesetting
\usepackage{booktabs}       % professional-quality tables
\usepackage{amsfonts}       % blackboard math symbols
\usepackage{nicefrac}       % compact symbols for 1/2, etc.
\usepackage{microtype}      % microtypography
\usepackage{xcolor}      % microtypography

\usepackage{amsmath}
\usepackage{stmaryrd}
\usepackage{caption}
\usepackage{subcaption}
\usepackage{graphicx}
\usepackage{subcaption}
\usepackage{multirow}
\usepackage{hypcap}
\usepackage{comment}
\usepackage{standalone}
\usepackage{enumitem}
\usepackage{xparse}
\usepackage{tikz}
\usepackage{tikz-3dplot} 
\usepackage{keyval}
\usepackage{ifthen}
\usepackage{float}
\usepackage[toc,page]{appendix}

\tdplotsetmaincoords{60}{125} % view angles
\tdplotsetrotatedcoords{0}{0}{0} 
\usetikzlibrary{positioning, fit, backgrounds}
\graphicspath{ {./} }
\newcommand*\samethanks[1][\value{footnote}]{\footnotemark[#1]}
\newcommand{\mc}[1]{\multicolumn{1}{l}{#1}}
\newcommand\eg{{\it{e.g.}}}
\newcommand\y{\mathbf y}
\newcommand\x{\mathbf x}
\newcommand\D{\mathbf D}
\newcommand\db{\mathbf d}
\newcommand\C{\mathbf C}

\newcommand\W{\mathbf W}
\newcommand\Real{\mathbb R}
\newcommand\vect{\text{vec}}
\newcommand\alphab{\boldsymbol \alpha}

\newcommand\hatx{\hat{\mathbf {{x}}}}
\newcommand\vs{{\vspace*{0cm}}}

\newcommand{\ts}{\textsuperscript}

\title{A Trainable Spectral-Spatial Sparse Coding Model for Hyperspectral Image Restoration}

\author{%
	Théo Bodrito\thanks{Equal contribution}~, Alexandre Zouaoui\samethanks~, Jocelyn Chanussot, and Julien Mairal\\
        Inria, Univ. Grenoble Alpes, CNRS, Grenoble INP, LJK, 38000 Grenoble, France \\
        \texttt{firstname.lastname@inria.fr}
}

\begin{document}

\maketitle

\begin{abstract}
	Hyperspectral imaging offers new perspectives for diverse applications, ranging
from the monitoring of the environment using airborne or satellite remote
sensing, precision farming, food safety, planetary exploration, or
astrophysics. Unfortunately, the spectral diversity of information comes at
the expense of various sources of degradation,  and the lack of accurate ground-truth ``clean'' hyperspectral signals
acquired on the spot makes restoration tasks challenging.
 In particular, training deep neural networks for restoration is difficult, in contrast
to traditional RGB imaging problems where deep models tend to shine. In this
paper, we advocate instead for a hybrid approach based on sparse coding
principles that retains the interpretability of classical techniques encoding
domain knowledge with handcrafted image priors, while allowing to train model
parameters end-to-end without massive amounts of data. We
show on various denoising benchmarks that our method is computationally efficient and  significantly outperforms the state of the art.
\footnote[1]{Code is available at \url{https://github.com/inria-thoth/T3SC}.}

\end{abstract}

\section{Introduction}

Hyperspectral imaging (HSI) enables measurements of the electromagnetic
spectrum of a scene on multiple bands (typically about a hundred or more), which
offers many perspectives over traditional color RGB imaging. For
instance, the high-dimensional information present in a single pixel is
sometimes sufficient to identify the signature of a particular material, which
is of course infeasible in the RGB domain. Not surprisingly, hyperspectral imaging is then of
utmost importance and has a huge number of scientific and technological applications such as
remote sensing \cite{bioucas-diasHyperspectralRemoteSensing2013, goetzThreeDecadesHyperspectral2009, manolakisHyperspectralImagingRemote2016},
quality evaluation of food products \cite{elmasryPrinciplesApplicationsHyperspectral2012,fengApplicationHyperspectralImaging2012,liuHyperspectralImagingTechnique2017},
medical imaging \cite{akbariHyperspectralImagingQuantitative2012, feiChapterHyperspectralImaging2020,luMedicalHyperspectralImaging2014},
agriculture and forestry \cite{adaoHyperspectralImagingReview2017, luRecentAdvancesHyperspectral2020,maheshHyperspectralImagingClassify2015},
microscopy imaging in biology \cite{gowenRecentApplicationsHyperspectral2015, studerCompressiveFluorescenceMicroscopy2012}, or exoplanet detection in astronomy~\cite{gonzalezSupervisedDetectionExoplanets2018}.
%and homeland security \cite{yuenIntroductionHyperspectralImaging2010, makkiSurveyLandmineDetection2017}.

Information contained in hyperspectral signals is much richer than in RGB
images, but the price to pay is the need to deal with complex degradations that may
arise from multiple sources, including sparse noise with specific
patterns (stripes), in addition to photon and thermal noise~\cite{kerekesHyperspectralImagingSystem2003,
	rastiNoiseReductionHyperspectral2018}.
As a consequence, HSI denoising is a crucial pre-processing step to enhance the
image quality before using data in downstream tasks such as semantic
segmentation or spectral unmixing~\cite{keshava2002spectral}.
A second issue is the lack of large-scale collection of ground-truth
high-quality signals and the large diversity of sensor types, which makes it particularly challenging to train machine
learning models for restoration such as convolutional neural
networks. To deal with the scarcity of ground-truth data, most successful
approaches typically encode strong prior
knowledge about data within the model architecture, which may be
low-rank representations of input patches~\cite{fanSpatialSpectralTotal2018,gongLowrankTensorDictionary2020,rastiAutomaticHyperspectralImage2017,wangHyperspectralImageMixed2020,zhaoHyperspectralImageDenoising2015a},
sparse coding~\cite{dantasHyperspectralImageDenoising2019,fuAdaptiveSpatialspectralDictionary2015,gongLowrankTensorDictionary2020},
or image self-similarities~\cite{maggioniNonlocalTransformDomainFilter2013,pengDecomposableNonlocalTensor2014,zhuangHyperspectralImageDenoising2017}, which have proven to be very powerful in the RGB domain~\cite{buadesNonlocalAlgorithmImage2005}.

In this paper, we propose a fully interpretable machine learning model for hyperspectral images that may be seen as a hybrid approach between deep learning
techniques, where parameters can be learned end to end with supervised data,
and classical methods that essentially rely on image priors.
Since designing an appropriate image prior by hand is very hard, our goal is to benefit
from deep learning principles (here, differentiable
programming~\cite{baydin2018automatic}) while encoding domain knowledge and physical rules
about hyperspectral data directly into the model architecture, which we believe is a key to develop robust approaches that do not require
massive amounts of training data.

More precisely, we
introduce a novel trainable spectral-spatial sparse coding model with two layers, which performs
the following operations:
(i) The first layer decomposes the spectrum measured at each pixel as a sparse
linear combination of a few elements from a learned dictionary, thus performing
a form of linear spectral unmixing per pixel, where dictionary elements can be seen
as basis elements for spectral responses of materials present in the scene.
(ii) The second layer builds upon the output of the first one, which is represented as a two-dimensional feature map, and sparsely encodes patches on a dictionary in order to take
into account spatial relationships between pixels within small receptive
fields.  To further reduce the number of parameters to learn and leverage
classical prior knowledge about spectral signals~\cite{wangHyperspectralImageMixed2020}, we also assume that the
dictionary elements admit a low-rank structure---that is, dictionary elements
are near separable in the space and spectrum domains, as detailed later.  Even
though dictionary learning has been originally introduced for unsupervised
learning~\cite{,mairalSparseModelingImage2014a,olshausen1996emergence},
we adopt an unrolled optimization procedure inspired by the LISTA
algorithm~\cite{gregorLearningFastApproximations2010a}, which has been very
successful in imaging problems for training sparse coding models from
supervised
data~\cite{lecouat_neurips,lecouatFullyTrainableInterpretable2020a,simonRethinkingCSCModel2019,xiongSMDSNetModelGuided2020}.

Our motivation for adopting a two-layer model is to provide a
shared architecture for different HSI sensors, which often involve a different
number of bands with different spectral responses. Our solution consists of learning sensor-specific dictionaries 
for the first layer, while the dictionary of second layer is shared across modalities. This
allows training simultaneously on several HSI signals, the first layer mapping
input data to a common space, before processing data by the second layer.

We experimentally evaluate our HSI model on standard denoising benchmarks,
showing a significant improvement over the state of the art (including deep
learning models and more traditional baselines), while being computationally very
efficient at test time.
Perhaps more important than pure quantitative results, we believe that our work
also draws interesting conclusions for machine learning.
First, by encoding prior knowledge within the model architecture directly, we obtain models
achieving excellent results with a relatively small number of parameters to
learn, a conclusion also shared
by~\cite{lecouat_neurips,lecouatFullyTrainableInterpretable2020a} for RGB imaging; nevertheless,
the effect is stronger in our work due to the scarcity of training data for HSI
denoising and the difficulty to train deep learning models for this task.
Second, we also show that interpretable architectures are useful:
our model architecture can adapt to different noise levels per band and 
modify the encoding function at test time in a principled manner, making it well suited
for solving blind denoising problems that are crucial for processing hyperspectral signals.

\section{Related Work on Hyperspectral Image Denoising}

\paragraph{Learning-free and low-rank approaches.}
Classical image denoising methods such as BM3D~\cite{dabovImageDenoisingSparse2007} may be
applied independently to each spectral band of HSI signals, but
such an approach fails to capture relations between channels;
Not surprisingly, multi-band techniques 
such as
BM4D~\cite{maggioniNonlocalTransformDomainFilter2013} have been shown to perform better for HSI, and other variants were subsequently proposed such as GLF
\cite{zhuangHyperspectralImageDenoising2017}.
Tensor-based methods such as LLRT~\cite{chang2017hyper} are able to exploit the underlying low-rank
structure of HSI signals~\cite{fanSpatialSpectralTotal2018,
	rastiAutomaticHyperspectralImage2017,zhangHyperspectralImageRestoration2014} and have shown particularly effective when combined with a non-local image
prior as in NGMeet~\cite{he2019non}.
Finally, other approaches adapt traditional image processing priors such as
total variation \cite{yuanHyperspectralImageDenoising2014, wangHybridTotalVariation2021},
or wavelet sparsity \cite{othmanNoiseReductionHyperspectral2006, rastiHyperspectralImageDenoising2014}
but they tend to perform worse than GLF, LLRT, or NGMeet, see~\cite{kong2020comprehensive} for a survey on denoising techniques for HSI.

\vs
\paragraph{Sparse coding models.}
Dictionary learning~\cite{olshausen1996emergence} is an
unsupervised learning technique consisting of representing a signal as a linear
combination of a few elements from a learned dictionary, which has shown
to be very effective for various image restoration
tasks~\cite{eladImageDenoisingSparse2006,mairal2007sparse}.
Several approaches have then combined dictionary learning and low-rank regularization.
For instance, 3D patches are represented as tensors in \cite{pengDecomposableNonlocalTensor2014} and are encoded by using spatial-spectral dictionaries \cite{tuckerMathematicalNotesThreemode1966}.
In \cite{zhaoHyperspectralImageDenoising2015a}, 2D patches are extracted from the band-vectorized representation of the 3D HSI data and sparsely encoded on a dictionary, while encouraging low-rank representations with a trace norm penalty on the reconstructed image.
The low-rank constraint can also be enforced by designing the dictionary as the result of the matrix multiplication between spatial and spectral dictionaries learned by principal component analysis as in~\cite{fuAdaptiveSpatialspectralDictionary2015}.
However, these methods typically compute sparse representations with an iterative optimization procedure, which may be computationally demanding at test time.

\vs
\paragraph{Deep learning.}
Like BM3D above, convolutional neural networks for grayscale image denoising
(\eg, DnCNN \cite{zhangGaussianDenoiserResidual2017}) may also be applied to
each spectral band, which is of course suboptimal.
Because deep neural networks have been highly successful for RGB images with often low computational inference cost, there
have been many attempts to design deep neural networks dedicated to HSI denoising.
For instance, to account for the large number of hyperspectral bands, several approaches based on convolutional neural networks are operating on sliding windows in the spectral domain,
\cite{maffeiSingleModelCNN2020a, shiHyperspectralImageDenoising2021,yuanHyperspectralImageDenoising2019},
which allows training models on signals with different number of spectral
bands, but the sliding window significantly increases the inference time.
More precisely, attention layers are used in~\cite{shiHyperspectralImageDenoising2021}, while more traditional CNNs are used in~\cite{maffeiSingleModelCNN2020a}, possibly with residual connections~\cite{yuanHyperspectralImageDenoising2019}.
Recently, an approach based on recurrent architecture was proposed in \cite{wei3DQuasirecurrentNeural2020} to process signals with an arbitrary number of bands, achieving impressive results for various denoising tasks.

\vs
\paragraph{Hybrid approaches.}
SMDS-Net \cite{xiongSMDSNetModelGuided2020} adopts a hybrid approach between sparse coding and deep learning models by adapting the RGB image restoration method of \cite{lecouatFullyTrainableInterpretable2020a} to HSI images.
The resulting pipeline however lacks interpretability:
SMDS-Net first denoises the input image with non-local means \cite{buadesNonlocalAlgorithmImage2005}, then performs subspace projection in the spectral domain using HySime\cite{bioucas-diasHyperspectralSubspaceIdentification2008}, before sparsely encoding 3D patches (cubes) with a trainable version of Tensor-based ISTA \cite{qiTenSRMultidimensionalTensor2016}.
Although this method reduces considerably the number of parameters in comparison to vanilla deep learning models, the spectral sliding windows approach lacks interpretability since the same denoising procedure is applied across different bands,
which may not suffer from the same level of noise. 
In contrast, we propose a much simpler sparse coding model, which is physically consistent with the nature of hyperspectral signals,
by introducing a novel differentiable low-rank sparse coding layer.

\section{A Trainable Spectral-Spatial Sparse Coding Model (T3SC)}

In this section, we introduce our trainable spectral-spatial sparse coding model dedicated to hyperspectral imaging,
and start by presenting some preliminaries on sparse coding.

\subsection{Background on Sparse Coding}

\paragraph{Image denoising with dictionary learning.}
A classical approach introduced by Elad and
Aharon~\cite{eladImageDenoisingSparse2006} for image denoising consists in considering the set of
small overlapping image patches (\eg, $8 \times 8$ pixels) from a noisy image, and compute a
sparse approximation of these patches onto a learned dictionary. The clean
estimates for each patch are then recombined to produce the full image.

Formally, let us consider a noisy image $\y$ in $\Real^{c \times h \times w}$ with $c$ channels and two spatial dimensions.
We denote by $\mathbf{y}_1, \mathbf{y}_2, \cdots \mathbf{y}_n$ the $n$
overlapping patches from $\y$ of size $c \times s \times s$, which we represent
as vectors in $\Real^m$ with $m=c s^2$.
Assuming that a dictionary $\D= [ \mathbf{d}_1, \cdots, \mathbf{d}_p ]$ in $\Real^{m \times p}$ is given---we will discuss later how to obtain a ``good'' dictionary---
each patch~$\y_i$ is processed by computing a sparse approximation:
\begin{equation}
	\label{eq:goal}
	\min_{\alphab_i \in \Real^p} \frac{1}{2} \| \mathbf{y}_i - \mathbf{D} {\alphab}_i \|^2 + \lambda \| {\alphab}_i \|_1,
\end{equation}
where $\| \cdot \|_1$ is the $l_1$-norm, which is known to induce sparsity in the problem solution
\cite{mairalSparseModelingImage2014a},
and $\alphab_i$ is the sparse code representing the patch~$\y_i$, while $\lambda$ controls the amount of regularization.
Note that the $\ell_0$-penalty, which counts the number of non-zero elements, could also be used, leading to a combinatorial
problem whose solution is typically approximated by a greedy algorithm.
After solving the~$n$ problems~(\ref{eq:goal}), each patch $\y_i$ admits a ``clean'' estimate
${\D} {\alphab}_i$. Because each pixel belongs to several patches, the full denoised image $\hatx$ is
obtained by averaging these estimates.

Finding a good dictionary can be achieved in various manners. In
classical dictionary learning algorithms, $\D$ is optimized such that
the sum of the loss functions~(\ref{eq:goal}) is as small as possible, see~\cite{mairalSparseModelingImage2014a} for a review.
Adapting the dictionary with supervision is also possible~\cite{Mairal_2012}, as discussed next.
\paragraph{Differentiable programming for sparse coding.}
The proximal gradient descent method called ISTA
\cite{figueiredoEMAlgorithmWaveletbased2003}
is a classical algorithm for solving the Lasso problem in Eq.~(\ref{eq:goal}),
which consists of the following iterations
\begin{equation}
	\label{eq:iteration}
	\mathbf{\alphab}_i^{(t + 1)} = S_{\lambda} \left[ {\alphab}_i^{(t)}  + \eta \D^\top \left( \y_i - \D \alphab_i^{(t)} \right) \right],
\end{equation}
where $\eta > 0$ is a step-size and $S_{\lambda}[u] =
	\text{sign}(u)\max(|u|-\lambda,0)$ is the soft-thresholding operator, which is applied pointwise to each entry of an input vector.

By noting that the above iteration can be seen as a sequence of affine
transformations interleaved with pointwise non-linearities $S_\lambda$, it is
then tempting to interpret $T$ iterations~(\ref{eq:iteration}) as a multilayer feed-forward
neural network with shared weights between the $T$ layers. Following such an insight, Gregor and LeCun have proposed the
LISTA algorithm~\cite{gregorLearningFastApproximations2010a}, where the parameters are learned such that
the sequence approximates well the solution of the sparse coding problem~(\ref{eq:goal}).

Interestingly, the LISTA algorithm can also be used to train dictionaries for
supervised learning tasks. This is the approach chosen in
\cite{lecouatFullyTrainableInterpretable2020a, simonRethinkingCSCModel2019} for image restoration, which considers the following iterations:
\begin{equation}
	\label{eq:iteration2}
	{\alphab}_i^{(t + 1)} = S_{\lambda} \left[ {\alphab}_i^{(t)}  + \C^\top \left( \y_i - \D \alphab_i^{(t)} \right) \right],
\end{equation}
which differs from~(\ref{eq:iteration}) with the presence of a matrix~$\C$ of
the same size as~$\D$.  Even if the choice $\C =\eta \D$ (which recovers ISTA)
is perfectly reasonable, using a different dictionary~$\C$ has empirically
shown to provide improvements in results
quality~\cite{lecouatFullyTrainableInterpretable2020a}, probably due to faster
convergence of the LISTA iterations.
Then, given some fixed parameters~$\C, \D$, a clean estimate $\W \alphab_i^{(T)}$ for each patch~$\y_i$ is obtained
by using a dictionary $\W$, where $T$ is the number of LISTA steps.
The reason for allowing a different
dictionary~$\W$ than~$\D$ is to correct the potential bias due
to~$\ell_1$-minimization.

Finally, the denoised image~$\hatx$ is
reconstructed by averaging the patch estimates:
\begin{equation}
	\label{eq:reconstruction2}
	\hatx = \frac{1}{m} \sum_{i=1}^{n} \mathbf{R}_i {\W} \mathbf{\alphab}_i^{(T)},
\end{equation}
where $\mathbf{R}_i$ is the linear operator that places the patch $\hat{\mathbf{x}}_i$ at position $i$ in the image, and we assume---by neglecting border effects for simplicity---that each pixel admits the same number~$m$ of estimates.

In contrast to classical restoration techniques based on dictionary learning, the LISTA point of view enables us to learn the model parameters~$\C, \D, \W$ in a supervised fashion. Given a training set of pairs of noisy/clean images, we
remark that the estimate $\hatx$ is obtained from a noisy image~$\y$ by a
sequence of operations that are differentiable almost everywhere, as typical neural networks with
rectified linear unit activation functions. A typical loss, which we optimize by stochastic gradient descent, is then
\begin{displaymath}
	\min_{\C, \D, \W, \lambda} {\mathbb E}_{\x, \y} \left[\| \hatx(\y) - \x \|^2 \right],
\end{displaymath}
where $(\x, \y)$ is a pair of clean/noisy images drawn from some training
distribution from which we can sample, and $\hatx(\y)$ is the clean estimate obtained
from~(\ref{eq:reconstruction2}), given the noisy image $\y$.

\subsection{A Trainable Low-Rank Sparse Coding Layer}

\begin{figure}
\centering
\input{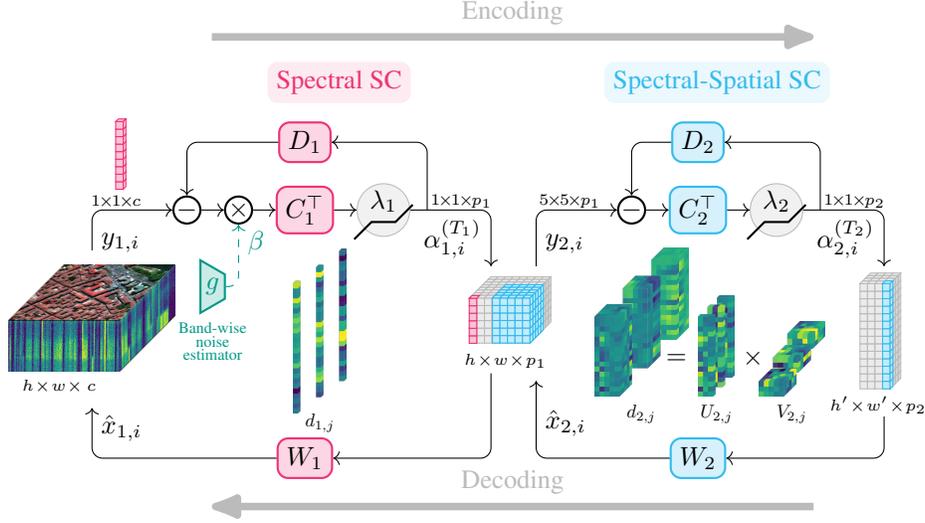}
\caption{Architecture of T3SC : we propose a two-layer sparse coding model which is end-to-end trainable. The first layer performs a sensor-specific spectral decomposition, while the second layer encodes both spectral and spatial information.}
\label{fig:architecture}
\end{figure}

We are now in shape to introduce a trainable layer encoding both sparsity and low-rank principles.

\vs
\paragraph{Spatial-Spectral Representation.}
As shown in \cite{chakrabartiStatisticsRealworldHyperspectral2011a,
	fuAdaptiveSpatialspectralDictionary2015},
HSI patches can be well reconstructed by using only a few basis elements obtained by 
principal component analysis.  The authors further
decompose these into a Cartesian product of separate spectral and spatial
dictionaries.
In this paper, we adopt a slightly different approach, where we consider
a single dictionary~$\D=[\db_1, \ldots, \db_p]$ in $\Real^{m \times p}$ as in the previous section with $m = c s^2$,
but each element may be seen as a matrix of size $c \times s^2$ with low-rank structure. More precisely, we enforce
the following representation
\begin{equation}
	\label{eq:decomp}
	\forall j \in 1,\ldots,p, \ \mathbf{d}_j = \vect\left( \mathbf{U}_j \times \mathbf{V}_j\right),
\end{equation}
where $ \mathbf{U}_j$ is  in $\mathbb{R}^{s^2 \times r} $, $ \mathbf{V}_j$ is  in $\mathbb{R}^{r \times c} $,
$r$ is the desired rank of the dictionary elements,
and $\vect(.)$ is the operator than flattens a matrix to a vector.
The hyperparameter $r$ is typically small with $r=1,2$ or $3$.  When $r=1$, the
dictionary elements are said to be separable in the spectral and spatial
domains, which we found to be a too stringent condition to achieve good reconstruction in practice.
%In our experiments, the choice $r=2$ or $3$ turned out to be a better choice.

The low-rank assumption allows us to build model with a reduced number of parameters, while encoding natural assumption about the data directly in the model architecture. Indeed, whereas a classical full-rank dictionary $\D$ admits $cs^2p$ parameters,
the decomposition~(\ref{eq:decomp}) yields dictionaries with $(s^2 + c)rp$ parameters only.
Matrices $ \mathbf{C}$ and $ \mathbf{W}$ are parametrized in a similar manner.

\vs
\paragraph{Convolutional variant and implementation tricks.}

Whereas traditional sparse coding reconstructs local signals (patches) independently according to the iterations~(\ref{eq:iteration2}), another variant called convolutional sparse coding (CSC) represents the whole image by a sparse linear combination of dictionary elements placed at
every possible location in the image~\cite{simonRethinkingCSCModel2019}. From a mathematical point of view, the reconstruction loss 
for computing the codes $\alphab_i$ given an input image~$\mathbf y$ becomes 
\begin{equation}
\label{eq:goal_csc}
\min_{\{\alphab_i \in \Real^p\}_{i=1,\ldots,n}} \frac{1}{2} \left\| \mathbf{y} - \frac{1}{m} \sum_{i=1}^{n} \mathbf{R}_i {\D} \mathbf{\alphab}_i \right\|^2 + \lambda \sum_{i=1}^n \| {\alphab}_i \|_1.
\end{equation}
An iterative approach for computing these codes can be obtained by a simple modification of~(\ref{eq:iteration2}) consisting of replacing the
quantity $\D \alphab_i^{(t)}$ by the $i$-th patch of the reconstructed image $\frac{1}{m} \sum_{i=1}^{n} \mathbf{R}_i {\D} \mathbf{\alphab}_i^{(t)}$. All of these operations can be efficiently implemented in standard deep learning frameworks, since the corresponding operations corresponds to a transposed convolution with $\D$, followed by convolution with $\C$, see~\cite{simonRethinkingCSCModel2019} for more details. In this paper, we experimented with the CSC variant~(\ref{eq:goal_csc}) and SC one~(\ref{eq:goal}), both with low-rank dictionaries, which were previously described. We observed that CSC was providing slightly better results and was thus adopted in our experiments.
Following~\cite{lecouatFullyTrainableInterpretable2020a}, another
implementation trick we use is to consider a different $\lambda$ parameter per
dictionary element, which slightly increases the number of parameters, while
allowing to learn with a weighted $\ell_1$-norm in~(\ref{eq:goal_csc}).

\subsection{The Two-Layer Sparse Coding Model with Sensor-Specific Layer}\label{subsec:vanilla}
One of the main challenge in hyperspectral imaging is to train a model that can
generalize to several types of sensors, which typically admit different number
of spectral bands.
Whereas learning a model that is tuned to a specific sensor is perfectly acceptable
in many contexts, it is often useful to learn a model that is able to generalize across
different types of HSI signals.
To alleviate this issue, several strategies have been adopted such as
(i) projecting signals onto a linear subspace of fixed dimension, with no guarantee that
representations within this subspace can be comparable between different signals, or (ii)
processing input data using a sliding window across the spectral domain.

In this paper, we address this issue by learning a two-layer model, presented in
Figure~\ref{fig:architecture}, where the first layer is tuned to a specific
sensor, whereas the second layer could be generic. Note that the second layer carries most of the model parameters (about $20\times$ more than in the first layer in our experiments). 
Formally, let us denote by
 $ {\alphab}$ in $\mathbb{R}^{p \times h \times w}$ the sparse encoding of
an input tensor $\mathbf{y}$ in $\mathbb{R}^{c \times h \times w}$ as previously described.
A sparse coding layer $ \Phi $ naturally yields an encoder and a decoder such that:
\begin{equation}
\Phi^{enc} : \mathbf{y}       \mapsto \mathbf{\alphab}, ~~~~\text{and}~~~~                                             
\Phi^{dec} : \mathbf{\alphab}  \mapsto \frac{1}{n} \sum_{i=1}^{n} \mathbf{R}_i \mathbf{W} \mathbf{\alphab}_i. 
\end{equation}
Given a noisy image $\y$, the denoising procedure described in the previous section with one layer can be written as 
$$ \mathbf{ \hat{x} }(\y) = \Phi^{dec} \circ \Phi^{enc}( \mathbf{y} ). $$
Then, a straightforward multilayer extension of the procedure may consist of stacking
several sparse coding layers $\Phi_1, \ldots, \Phi_L$ together to form a multilayer sparse coding denoising model:
$$ \mathbf{ \hat{x} }(\y) = \Phi_1^{dec} \circ \cdots \circ \Phi_L^{dec} \circ \Phi_L^{enc} \circ \cdots \circ \Phi_1^{enc}( \mathbf{ y } ).$$
The model we propose is composed of two layers, as shown in Figure~\ref{fig:architecture}.
The first layer encodes spectrally the input HSI image, meaning that it operates on $1 \times 1$ patches, whereas the second layer encodes both spectrally and spatially the output of the first layer.

\subsection{Noise Adaptive Sparse Coding}\label{subsec:band}
An advantage of using a model based on a sparse coding
objective~(\ref{eq:goal}) is to give the ability to encode domain knowledge within
the model architecture. 
For instance, the Lasso problem~(\ref{eq:goal}) seen from a maximization a posteriori estimator
implicitly assumes that the noise is i.i.d.
If the noise variance is different on each spectral band, a natural modification to the 
model is to introduce weights and use a weighted-$\ell_2$ data fitting term (which could be applied as well to the CSC model of~(\ref{eq:goal_csc})):
\begin{equation}
\label{eq:goal2}
\min_{\alphab_i \in \Real^p} \frac{1}{2} \sum_{j=1}^c \beta_j \| \mathbf{M}_j ( \mathbf{y}_i - \mathbf{D} {\alphab}_i) \|^2 + \lambda \| {\alphab}_i \|_1,
\end{equation}
where $\mathbf{M}_j$ is a linear operator that extracts band $j$ from a given HSI signal. 
From a probabilistic point of view, if $\sigma_j^2$ denotes the variance of the noise for band $j$, we may choose the corresponding weight~$\beta_j$ to be proportional to $1/\sigma_j^2$.
Yet, estimating accurately $\sigma_j^2$ is not always easy, and we have
found it more effective to simply learn a parametric function $\beta_j = g
(\mathbf{M}_j  \mathbf{y})$---here, a very simple CNN with three layers, see supplementary material for details---which is applied independently to each band.
It is then easy to modify the LISTA iterations accordingly to take into account these weights, and learn the model parameters jointly with those of the parametric function~$g$.

\subsection{Self-Supervised Learning: Blind-Band Denoising with No Ground Truth Data}\label{subsec:ssl}
Even though acquiring limited ground truth data for a specific sensor is often
feasible, it is also interesting to be able to train models with no ground
truth at all, \eg, for processing images without physical access to the sensor. 
In such an unsupervised setting, deep neural networks
are typically trained for RGB images by using blind-spot denoising
techniques~\cite{laine2019high}, consisting of predicting pixel values given
their context. Here, 
we propose a much simpler approach exploiting the spectral redundancy between channels.
More precisely, each time we draw an image for training, we randomly mask one band (ore more), and
train the model to reconstruct the missing band from the available ones.
Formally, the training objective becomes
\begin{equation}
	\label{eq:goal4}
   \min_{\C,\D,\W,\lambda}  {\mathbb E}_{\x,\y,S} \left[  \sum_{j \notin S} \| \mathbf{M}_j (\hatx_S(\y) - \y )\|^2 \right],
\end{equation}
where $S$ is the set of bands that are visible for computing the sparse codes $\alphab_i$, leading to a reconstructed image that we denote by
$\hatx_S(\y)$. Formally, it would mean considering the objective~(\ref{eq:goal2}), but replacing the sum $\sum_{j=1}^c$ by $\sum_{j \in S}$. 
This is in spirit similar to blind-spot denoising, except that bands are masked instead of pixels, making the resulting implementation much simpler.

\section{Experiments}

We now present various experiments to demonstrate the effectiveness of our
approach for HSI denoising, but first, we discuss the difficulty of defining
the state of the art in this field.  We believe indeed that it is not always
easy to compare learning-free from approaches based on supervised learning.
These two classes of approaches have very different
requirements/characteristics, making one class more relevant than the other one
in some scenarios, and less in others. Table~\ref{table:comp} summarizes their
characteristics, displaying advantages and drawbacks of both approaches.
\begin{table}
   \caption{Simplified comparison between learning-free and learning-based approaches.}\label{table:comp}
   \vspace*{0.05cm}
   \centering
\begin{tabular}{llllll}
   & \emph{Data req.} & \emph{training} & \emph{inference} &  \emph{adapt. to new data} & \emph{complex noise} \\
                  \hline
   {\bfseries learning-free}  & no req. & no training & slow & easy & poor \\
                  \hline
   {\bfseries learning-based}  & clean data & slow & fast & complicated & good perf. \\
                  \hline

\end{tabular}
\end{table}

\vs
\paragraph{Benchmarked models.}
Keeping in mind the previous dichotomy, we choose to compare our method to traditional methods such as 
bandwise BM3D \cite{dabovImageDenoisingSparse2007} (implementation based on \cite{makinenExactTransformdomainNoise2019, makinenCollaborativeFilteringCorrelated2020}), 
BM4D \cite{maggioniNonlocalTransformDomainFilter2013},
GLF \cite{zhuangHyperspectralImageDenoising2017},
LLRT \cite{chang2017hyper},
NGMeet \cite{he2019non}.
We also included deep learning models in our benchmark such as  
HSID-CNN \cite{yuanHyperspectralImageDenoising2019}, 
HSI-SDeCNN \cite{maffeiSingleModelCNN2020a}
3D-ADNet \cite{shiHyperspectralImageDenoising2021},
SMDS-Net \cite{xiongSMDSNetModelGuided2020} and
QRNN3D \cite{wei3DQuasirecurrentNeural2020}.
Results of HSID-CNN, HSI-SDeCNN and 3D-ADNet on Washington DC Mall (available in the Appendix)
are taken directly from the corresponding papers, as the train/test split is the same.
Otherwise, the results were obtained by running the code obtained directly from the authors, except for SMDS-Net, where our implementation turned out to be slightly more effective.
Note that the same architecture for our model was used in all our experiments (see Appendix).

\paragraph{Datasets.}
We evaluate our approach on two datasets with significantly different properties.
\begin{itemize}[leftmargin=*,itemsep=0pt,parsep=0pt,topsep=0pt]
   \item \textit{ICVL} \cite{arad2016sparse}  consists of 204 images of size $1392 \times 1300$ with 31 bands. We used 100 images for training and 50 for testing as in
      \cite{wei3DQuasirecurrentNeural2020} but with a different train/test
      split ensuring that similar images---\eg, picture from the same scene---are not used twice.  
   \item \textit{Washington DC Mall} is perhaps the most widely used dataset\footnote{\url{https://engineering.purdue.edu/~biehl/MultiSpec/hyperspectral.html}} for HSI denoising and consists of 
    a high-quality image of size $1280 \times 307$ with 191 bands.  
Following~\cite{shiHyperspectralImageDenoising2021}, we split the image into 
two sub-images of size $600 \times 307$ and $480 \times 307$ for training and 
one sub-image of size $200 \times 200$ for testing. Even though the test image
      does not overlap with train images, they nevertheless
      share common characteristics. Interestingly, the amount of training data is very limited here.
\end{itemize}
Specific experiments were also conducted with the datasets APEX~\cite{itten2008apex}, Pavia\footnote{\url{http://www.ehu.eus/ccwintco/index.php?title=Hyperspectral_Remote_Sensing_Scenes}}, Urban\cite{urbandataset} and CAVE~\cite{CAVE}, which appear in the supplementary material.

\vs
\paragraph{Normalization.}
Before denoising, HSI images are normalized to $[0,1]$.
For remote sensing datasets, we pre-compute the 2\ts{nd} and 98\ts{th} percentiles for each band, on the whole the training set.
Then, normalization is performed on train and test images by clipping each band between those percentiles before applying bandwise min-max normalization, similar to \cite{audebertDeepLearningClassification2019a, maffeiSingleModelCNN2020a}.
For the close-range dataset ICVL, we simply apply global min-max normalization as in \cite{xiongSMDSNetModelGuided2020, wei3DQuasirecurrentNeural2020}.

\vs
\paragraph{Noise patterns.}
We evaluate our model against different types of synthetic noise:
\begin{itemize}[leftmargin=*,itemsep=0pt,parsep=0pt,topsep=0pt]
	\item \textit{i.i.d Gaussian noise with known variance} $\sigma^2$, which is the same on all bands.
	\item \textit{Gaussian noise with unknown band-dependent variance}: We consider Gaussian noise with different standard deviation $\sigma_j$ for each band, which is uniformly drawn in a fixed interval. These standard deviations change from an image to the other and are unknown at test time.
	\item \textit{Noise with spectrally correlated variance}: We consider Gaussian noise with standard deviation~$\sigma_j$ varying continuously across bands, following a Gaussian curve, see details in the appendix.
	\item \textit{Stripes noise} : similar to \cite{wei3DQuasirecurrentNeural2020}, we applied additive stripes noise to 33\% of bands.
	      In those bands, 10-15\% of columns are affected, meaning a value uniformly sampled in the interval $[-0.25, 0.25]$ is added to them.
	      Moreover, all bands are disturbed by Gaussian noise with noise intensity $\sigma=25$.
\end{itemize}

\vs
\paragraph{Metrics.}
In order to assess the performances the previous methods, we used five different indexes widely used for HSI restoration, namely
(i) Mean Peak Signal-to-Noise Ratio (MPSNR), which is the classical PSNR metric
averaged across bands; (ii) Mean Structural Similarity Index Measurement
(MSSIM), which is based on the SSIM metric \cite{zhou2004image}; 
 (iii) Mean Feature Similarity Index Measurement (MFSIM) introduced in \cite{zhang2011fsim};
 (iv) Mean ERGAS~\cite{du2007performance}, and
 (v) Mean Spectral Angle Map (MSAM)~\cite{alparone2007comparison}.
We use MPSNR and MSSIM in the main paper and report the other metrics in the appendix.

\vs
\paragraph{Implementation details.}
We trained our network by minimizing the MSE between the groundtruth and restored images.
For ICVL, we follow the training procedure described in \cite{wei3DQuasirecurrentNeural2020}: we first center crop training images to size $1024 \times 1024$, 
then we extract patches of size $64 \times 64$ at scales 1:1, 1:2, and 1:4, with stride 64, 32 and 32 respectively.
The number of extracted patches for ICVL amounts to 52962.
For Washington DC Mall, we do not crop training images and the patches are extracted with stride 16, 8 and 8, for a total of 1650 patches.
One epoch in Washington DC Mall corresponds to 10 iterations on the training dataset.
Basic data augmentation schemes such as $90^\circ$ rotations and vertical/horizontal flipping are performed.
Code and additional details about optimization, implementation, computational resources, are provided in the supplementary material.
As reported in Table~\ref{table:unrolling}, augmenting the number unrolled iterations improves the denoising performances at the expense of inference time.
Since the Spectral-Spatial SC layer is the most time-consuming, the number of unrolled iterations chosen for the first and second layers are 12 and 5 respectively.

\vs
\paragraph{Quantitative results on synthetic noise.}
We present in Table~\ref{table:icvl} the results obtained on the ICVL dataset (results on DCMall are presented in the appendix). 
Our method uses the vanilla model of Section~\ref{subsec:vanilla} for the experiments with constant $\sigma$ or correlated noise. 
For the blind denoising experiment with band-dependent $\sigma$ or for the stripe noise experiment, we use the variant of Section~\ref{subsec:band},
which is designed to deal with unknown noise level per channel. The method
``T3SC-SSL'' implements the self-supervised learning approach of Section~\ref{subsec:ssl}, which does not
rely on ground-truth data.
\begin{itemize}[leftmargin=*,itemsep=0pt,parsep=0pt,topsep=0pt]
   \item Our supervised approach achieves state-of-the-art results (or is close to the best performing baseline) on all settings. GLF performs remarkably well given that this baseline is learning-free. 
   \item Our self-supervised method achieves a relatively good performance under i.i.d. Gaussian noise, but does not perform as well under more complex noise.
     This is a limitation of the approach which is perhaps expected and overcoming this limitation would require designing a different self-supervised learning scheme; this is an interesting problem, which is beyond the scope of this paper. 
\end{itemize}

\begin{figure}[h!]
  \input{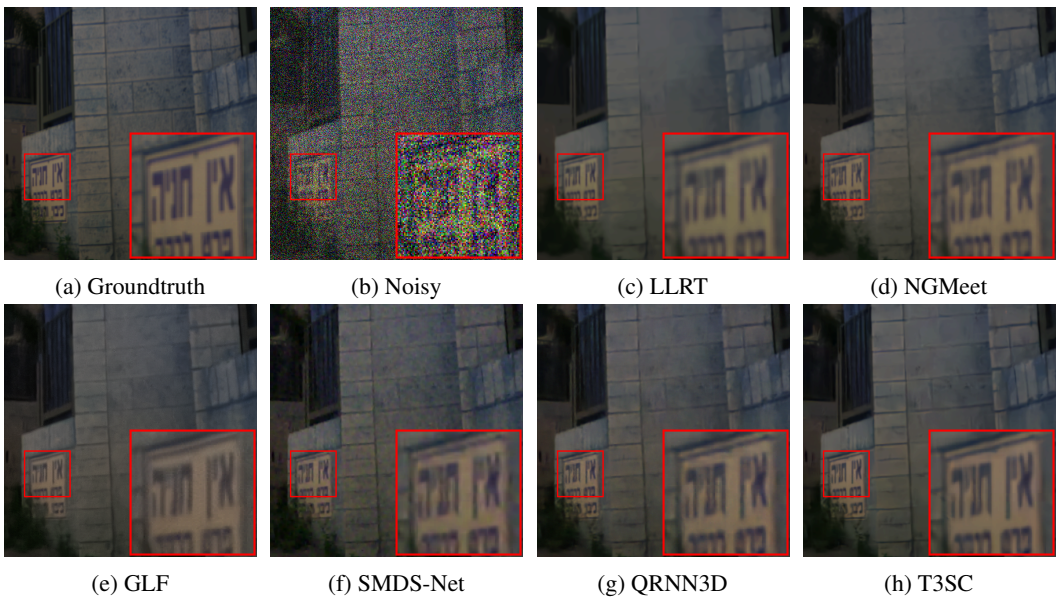}
        \caption{Denoising results with Gaussian noise $\sigma=25$ on ICVL with bands 9, 15, 28.}\label{fig:icvl}
\end{figure}

A visual result on ICVL is shown in Figure~\ref{fig:icvl} for stripes noise.
Inference times are provided in Table~\ref{table:speed}, showing that our approach is computationally efficient.

\begin{figure}[h!]
  \input{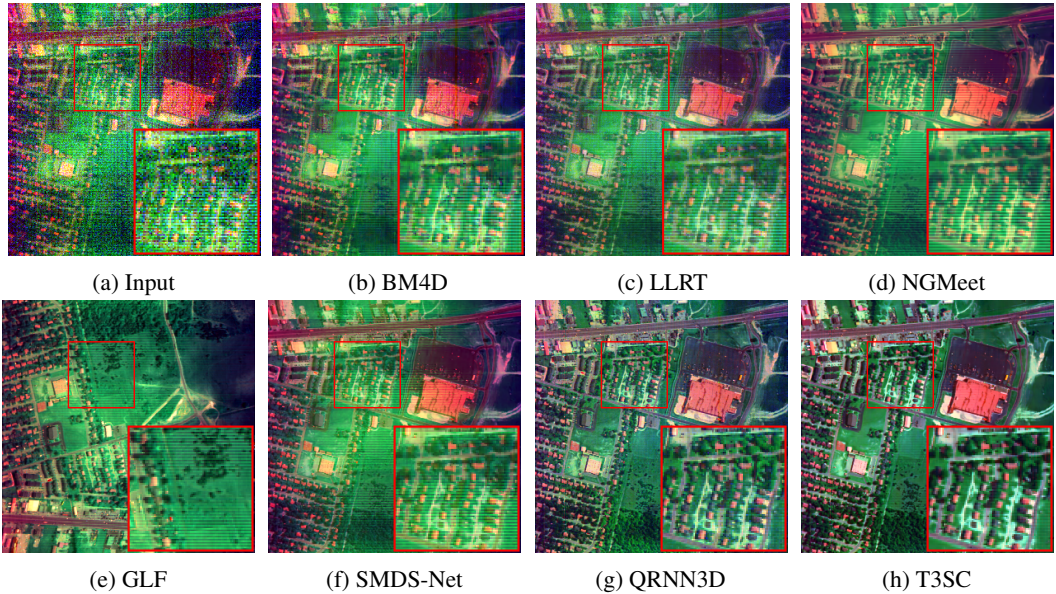}
        \caption{Visual result on a real HSI denoising experiment on Urban dataset with bands 1, 108, 208.}\label{fig:urban}
\end{figure}

\vs
\paragraph{Results on real noise.}
We also conducted a denoising experiment on the Urban dataset, reporting a visual result in Figure~\ref{fig:urban}. 
Deep models were pre-trained on the APEX dataset, which has the same number of channels as Urban (even though the sensors are different), with band-dependent noise with $\sigma \in [0-55] $.
Please note that for this experiment we did not use Noise Adaptive Sparse Coding\ref{subsec:band} for T3SC, as it is highly dependent on the type of sensor used for training.
We show that learning-based models trained on synthetic noise are able to transfer to real data.

\vs
\paragraph{Comments on the additional results presented in the appendix.}
The appendix also contains (i) results on the DCMall dataset including additional baselines mentioned above;
(ii) error bars for parts of our experimental results in order to assess their statistical significance; (iii) an experiment when learning simultaneously on several datasets with different types of sensors showing that the second layer can be generic and effective at the same time;
(iv) additional visual results; (v)  
various ablation studies to illustrate the importance of different components of our method.
\begin{table}[t]
   \captionof{table}{Denoising performance on ICVL with various types of noise patterns. The first four rows correspond to i.i.d. Gaussian noise with fixed~$\sigma$ per band. The next three rows corresponds to a noise level that depends on the band, taken uniformly on small interval. This is a blind-noise experiment since at test time, the noise level is unknown. The last two rows correspond to the scenarios with correlated $\sigma$ across bands, and with stripe noise, respectively. See main text for details.  } \label{table:icvl}
	\def\arraystretch{1.2}
\resizebox{\textwidth}{!}{
	\begin{tabular}{c c c c c c c c c c c c}
		\toprule
		$\hspace{1pt}\sigma$ \hspace{1pt}            & Metrics    & Noisy       & BM3D        & BM4D        & GLF         & LLRT        & NGMeet                  & SMDS        & QRNN3D                  & T3SC                    & T3SC-SSL                \\ [0.5ex]
		\hline\hline

		\multirow{2}{*}{\hspace{5pt}  5            } & \mc{MPSNR} & \mc{34.47}  & \mc{46.17}  & \mc{48.85}  & \mc{51.25}  & \mc{51.86}  & \mc{\textbf{52.74}}     & \mc{50.91}  & \mc{48.80}              & \mc{\underline{52.62}}  & \mc{51.42}              \\
		                                             & \mc{MSSIM} & \mc{0.7618} & \mc{0.9843} & \mc{0.9916} & \mc{0.9949} & \mc{0.9951} & \mc{\textbf{0.9960}}    & \mc{0.9944} & \mc{0.9918}             & \mc{\underline{0.9959}} & \mc{0.9952}             \\
		\hline
		\multirow{2}{*}{\hspace{5pt} 25 }            & \mc{MPSNR} & \mc{21.44}  & \mc{37.86}  & \mc{39.89}  & \mc{43.16}  & \mc{43.43}  & \mc{\underline{44.74}}  & \mc{42.83}  & \mc{44.20}              & \mc{\textbf{45.38}}     & \mc{44.73}              \\
		                                             & \mc{MSSIM} & \mc{0.1548} & \mc{0.9269} & \mc{0.9510} & \mc{0.9695} & \mc{0.9746} & \mc{\underline{0.9796}} & \mc{0.9700} & \mc{0.9782}             & \mc{\textbf{0.9825}}    & \mc{0.9805}             \\
		\hline
		\multirow{2}{*}{\hspace{5pt} 50 }            & \mc{MPSNR} & \mc{16.03}  & \mc{34.22}  & \mc{34.22}  & \mc{39.26}  & \mc{39.69}  & \mc{41.08}              & \mc{39.25}  & \mc{\underline{41.67}}  & \mc{\textbf{42.16}}     & \mc{41.62}              \\
		                                             & \mc{MSSIM} & \mc{0.0502} & \mc{0.8654} & \mc{0.8654} & \mc{0.9197} & \mc{0.9504} & \mc{0.9602}             & \mc{0.9382} & \mc{\underline{0.9655}} & \mc{\textbf{0.9677}}    & \mc{0.9646}             \\
		\hline
		\multirow{2}{*}{\hspace{5pt} 100 }           & \mc{MPSNR} & \mc{10.85}  & \mc{30.43}  & \mc{32.47}  & \mc{34.79}  & \mc{36.39}  & \mc{37.55}              & \mc{35.64}  & \mc{37.19}              & \mc{\textbf{38.99}}     & \mc{\underline{38.50}}  \\
		                                             & \mc{MSSIM} & \mc{0.0144} & \mc{0.7557} & \mc{0.8155} & \mc{0.7982} & \mc{0.9182} & \mc{0.9311}             & \mc{0.8815} & \mc{0.9140}             & \mc{\textbf{0.9439}}    & \mc{\underline{0.9394}} \\
		\hline
		\hline
		\multirow{2}{*}{\hspace{5pt} [0-15] }        & \mc{MPSNR} & \mc{33.89}  & \mc{45.81}  & \mc{45.35}  & \mc{50.57}  & \mc{48.50}  & \mc{41.67}              & \mc{48.23}  & \mc{\underline{52.07}}  & \mc{\textbf{53.31}}     & \mc{51.26}              \\
		                                             & \mc{MSSIM} & \mc{0.6386} & \mc{0.9767} & \mc{0.9735} & \mc{0.9948} & \mc{0.9899} & \mc{0.9078}             & \mc{0.9900} & \mc{\underline{0.9957}} & \mc{\textbf{0.9967}}    & \mc{0.9955}             \\
		\hline
		\multirow{2}{*}{\hspace{5pt}[0-55] }         & \mc{MPSNR} & \mc{23.36}  & \mc{39.06}  & \mc{38.43}  & \mc{44.22}  & \mc{41.13}  & \mc{32.94}              & \mc{41.76}  & \mc{\underline{47.13}}  & \mc{\textbf{48.64}}     & \mc{46.82}              \\
		                                             & \mc{MSSIM} & \mc{0.2601} & \mc{0.9231} & \mc{0.9074} & \mc{0.9818} & \mc{0.9580} & \mc{0.7565}             & \mc{0.9620} & \mc{\underline{0.9884}} & \mc{\textbf{0.9911}}    & \mc{0.9882}             \\
		\hline
		\multirow{2}{*}{\hspace{5pt}[0-95] }         & \mc{MPSNR} & \mc{19.06}  & \mc{36.17}  & \mc{35.55}  & \mc{41.43}  & \mc{38.44}  & \mc{29.40}              & \mc{38.94}  & \mc{43.98}              & \mc{\textbf{46.30}}     & \mc{\underline{44.75}}  \\
		                                             & \mc{MSSIM} & \mc{0.1614} & \mc{0.8760} & \mc{0.8540} & \mc{0.9674} & \mc{0.9354} & \mc{0.6609}             & \mc{0.9357} & \mc{0.9753}             & \mc{\textbf{0.9859}}    & \mc{\underline{0.9822}} \\
		\hline
		\hline

		\multirow{2}{*}{\hspace{5pt}Corr.}           & \mc{MPSNR} & \mc{28.85}  & \mc{42.73}  & \mc{42.13}  & \mc{47.05}  & \mc{45.76}  & \mc{38.06}              & \mc{45.98}  & \mc{\underline{48.90}}  & \mc{\textbf{49.89}}     & \mc{48.78}              \\
		                                             & \mc{MSSIM} & \mc{0.4740} & \mc{0.9599} & \mc{0.9070} & \mc{0.9881} & \mc{0.9824} & \mc{0.8536}             & \mc{0.9835} & \mc{\underline{0.9911}} & \mc{\textbf{0.9923}}    & \mc{\underline{0.9911}} \\
		\hline
		\hline

		\multirow{2}{*}{\hspace{5pt}Strip.}          & \mc{MPSNR} & \mc{21.20}  & \mc{34.88}  & \mc{37.70}  & \mc{42.06}  & \mc{39.38}  & \mc{39.78}              & \mc{41.98}  & \mc{\underline{44.60}}  & \mc{\textbf{44.74}}     & \mc{43.80}              \\
		                                             & \mc{MSSIM} & \mc{0.1508} & \mc{0.8641} & \mc{0.9198} & \mc{0.9628} & \mc{0.9258} & \mc{0.9333}             & \mc{0.9655} & \mc{\textbf{0.9806}}    & \mc{\underline{0.9805}} & \mc{0.9773}             \\
		\bottomrule
	\end{tabular}
}

\end{table}
\begin{table}[t]
   \centering
  \captionof{table}{Inference time per image on ICVL with $\sigma=50$; SMDS, QRNN3D and T3SC are using a V100 GPU; BM4D, GLF, LLRT and NGMeet are using an Intel(R) Xeon(R) CPU E5-1630 v4 @ 3.70GHz. Note that unlike GLF, NGMeet, and LRRT, learning-based approaches such as QRNN3D and our approach require a training procedure, which may be conducted offline. The cost of such a training step was about 13.5 hours for our method and 19 hours for QRNN3D on a V100 GPU. }\label{table:speed}
   
\footnotesize
\def\arraystretch{1.2}
\resizebox{\textwidth}{!}{
\begin{tabular}{ c c c c c c c c c c}
	\toprule
	                        & BM3D        & BM4D        & GLF         & LLRT & NGMeet      & SMDS      & QRNN3D             & T3SC     & T3SC-SSL   \\ %[0.5ex]
	\hline\hline
  \mc{Inference time (s)} & \mc{1677} & \mc{2382} & \mc{5565} & \mc{24384}     & \mc{2686} & \mc{74.3} & \mc{\textbf{3.6}} & \mc{\underline{5.8}} & \mc{54.2} \\
	\bottomrule
\end{tabular}
}

\end{table}
\begin{table}[t]
  \centering
   \captionof{table}{Impact of the number of unrolled iterations per layer on denoising performances and inference time. This experiment was carried out on ICVL with $\sigma=50$.} \label{table:unrolling}
	\def\arraystretch{1.2}
\resizebox{0.65\textwidth}{!}{
	\begin{tabular}{c | c c c c }
		\toprule
		Unrolled iterations per layer & 1     & 2     & 5     & 12    \\
		\hline
		MPSNR                         & 40.16 & 41.48 & 42.15 & 42.45 \\
		Inference time (s)            & 0.38  & 1.44  & 5.27  & 14.91 \\
		\bottomrule
	\end{tabular}
}

\end{table}

\subsection*{Broader Impact}
Our paper addresses the problem of denoising the signal, which is a key
pre-processing step before using hyperspectral signals in concrete
applications. As such, it is necessarily subject to dual use.
For instance, HSI may be used for environmental monitoring, forestry, yield
estimation in agriculture, natural disaster management planning, astronomy,
archaeology, and medicine.
Yet, HSI is also used by the petroleum industry for finding new oil fields,
and has obvious military applications for surveillance.
We believe the potential benefits of HSI for society are large enough to
outweigh the potential harm. Nevertheless, we are planning to implement
appropriate dissemination strategies to mitigate the risk of misuse for this
work (notably with restrictive software licenses), while targeting a gold standard regarding the scientific reproducibility of our results.

\subsection*{Acknowledgments and Funding}
This project was supported by the ERC grant number 714381 (SOLARIS project) and by ANR 3IA MIAI@Grenoble Alpes (ANR-19-P3IA-0003).
This work was granted access to the HPC resources of IDRIS under the allocation 2021-[AD011012382] made by GENCI.

\small
\bibliographystyle{abbrv}
\bibliography{main}

\newpage
\begin{center}
  \huge Supplementary Material \\
\end{center}

\appendix
\section{Implementation details}
In this section, we provide additional implementation details, which
are useful to reproduce our experiments (note that the code is also provided).
\paragraph{Noise with spectrally correlated variance.}
For each band $i \in \llbracket 0, c -1 \rrbracket$, the standard deviation of the Gaussian noise is defined as :
$$
\sigma_i = \beta \ exp \left[ - \frac{1}{4 \eta^2} \left(\frac{i}{c} - \frac{1}{2}\right)^2  \right]
$$
with $\beta=23.08$ and $\eta=0.157$.

\paragraph{Preprocessing.}
A basic centering step is used for each input patch of our model.
More precisely, for the first layer, each band of the input hyperspectral image is
centered independently prior to patches extraction, and means are added
back after decoding.
For the second layer, patches are centered independently for each band (and similarly, the means are added back after decoding).

\paragraph{Code and patch sizes}
The hyperparameters of our model are presented in Table~\ref{table:model_arch}.

\begin{table}[H]
	\centering
	\def\arraystretch{1.2}
\resizebox{0.7 \textwidth}{!}{
	\begin{tabular}{c | c c c c  }
		\toprule
		Layer               & Patches size & Code size & Unrolled iterations & Rank \\ [0.5ex]
		\hline\hline
		Spectral SC         & $1 \times 1$ & 64        & 12                  & 1    \\
		Spectral-Spatial SC & $5 \times 5$ & 1024      & 5                   & 3    \\
		\bottomrule
	\end{tabular}
}

	\captionof{table}{Architecture of our model}
	\label{table:model_arch}
\end{table}

Table~\ref{table:multilayer} shows that the combination of both layers is more performant than each layer independently.

\paragraph{Initialization}
All parameters are initialized with He initialization \cite{he2015delving}.

\paragraph{Blocks inference}
In order to apply our model to large images, we split them into blocks of size $256 \times 256$ with an overlap of 6 pixels.
Each block is denoised independently.
The output image is obtained by aggregating the denoised blocks.
Pixels comprised in several blocks are averaged.

\paragraph{Weights estimator}
For complex noise such as Gaussian noise with band-dependent variance or stripes noise, our model uses a CNN to estimate the weights $\beta_j$ associated with each band.
The CNN operates on centered patches of size $56 \times 56$ both during training (random crops) and inference (blocks inference), and its architecture is described in Table \ref{table:cnn_arch}.

\begin{table}[H]
	\centering
	\def\arraystretch{1.2}
\resizebox{0.7 \textwidth}{!}{
	\begin{tabular}{c | c c c c  }
		\toprule
		Layer            & Kernel size & Stride & \#filters & Output size               \\ [0.5ex]
		\hline\hline
		Inputs           &             &        &           & $1 \times 56 \times 56$   \\
		Conv2D + ReLU    & 5           & 2      & 64        & $ 64 \times 26 \times 26$ \\
		MaxPooling2D     &             &        &           & $ 64 \times 13 \times 13$ \\
		Conv2D + ReLU    & 3           & 2      & 128       & $ 128 \times 6 \times 6$  \\
		MaxPooling2D     &             &        &           & $ 128 \times 3 \times 3$  \\
		Conv2D + Sigmoid & 3           & 1      & 1         & $ 1 \times 1 \times 1$    \\
		\bottomrule
	\end{tabular}
}

	\captionof{table}{CNN architecture for estimating $\beta_j$}
	\label{table:cnn_arch}
\end{table}
The ablation study presented in Table \ref{table:beta} shows that this extension improves performances substantially for complex noise.

\paragraph{Optimization}
Our models are trained with batch size of 16 for 60 epochs.
We use the Adam optimizer,
the initial learning rate is $3 \times 10^{-4}$, and is divided by two at epoch 30 and 45.

\paragraph{Self-Supervised Learning}
During the training, $n$ bands randomly selected are masked simultaneously, and reconstructed from the available ones.
The MSE loss is applied on the masked bands only.
For testing, the $n$ masked bands are evenly distributed along the spectral dimension.
All bands are reconstructed after $ \lceil c / n \rceil $ iterations, where $c$ denotes the total number of bands and $\lceil \cdot \rceil$ is the ceiling operator.
We used $n=4$ for ICVL and $n=16$ for Washington DC Mall.

For the SSL setting to be realistic, the noise is added to the clean image before patches are extracted.
Otherwise, the model would have indirect access to the groundtruth by seeing the same patch with different noise realizations.
As a consequence, the denoising task is much harder on complex noise when data is limited, as shown in Table~\ref{table:dcmall}.

\section{Additional Quantitative Results.}
\paragraph{Washington DC Mall dataset.} Results for this dataset are presented in Table~\ref{table:dcmall}. Additional baselines are presented in Table~\ref{table:dcmallext}.
The conclusions are similar to those already drawn in the main paper.

\begin{table}[H]
	\centering
	\captionof{table}{Denoising performances on Washington DC Mall.}\label{table:dcmall}
	\def\arraystretch{1.2}
\resizebox{\textwidth}{!}{
	\begin{tabular}{c c c c c c c c c c c c c c c c}
		\toprule
		$\hspace{1pt}\sigma$ \hspace{1pt}     & Metrics & Noisy  & BM3D   & BM4D   & GLF                & LLRT   & NGMeet             & SMDS               & QRNN3D             & T3SC               & T3SC-SSL           \\ [0.5ex]
		\hline\hline
		\multirow{2}{*}{\hspace{5pt} 5  }     & MPSNR   & 34.31  & 35.10  & 41.13  & 39.57              & 41.83  & 37.57              & 42.83              & \underline{43.42}  & \textbf{43.85}     & 42.56              \\
		                                      & MSSIM   & 0.9821 & 0.9875 & 0.9962 & 0.9953             & 0.9968 & 0.9928             & 0.9971             & \underline{0.9973} & \textbf{0.9978}    & 0.9967             \\
		\hline
		\multirow{2}{*}{\hspace{5pt} 25 }     & MPSNR   & 20.70  & 24.51  & 31.08  & 35.25              & 34.95  & 35.38              & 35.64              & 35.04              & \textbf{36.74}     & \underline{35.92}  \\
		                                      & MSSIM   & 0.7688 & 0.8859 & 0.9690 & 0.9883             & 0.9863 & 0.9886             & 0.9889             & 0.9864             & \textbf{0.9912}    & \underline{0.9894} \\
		\hline
		\multirow{2}{*}{\hspace{5pt} 50 }     & MPSNR   & 15.25  & 20.80  & 26.69  & 31.77              & 30.94  & 31.88              & 31.76              & 31.72              & \textbf{33.12}     & \underline{31.96}  \\
		                                      & MSSIM   & 0.5314 & 0.7508 & 0.9220 & 0.9761             & 0.9704 & 0.9759             & \underline{0.9765} & 0.9741             & \textbf{0.9819}    & 0.9762             \\
		\hline
		\multirow{2}{*}{\hspace{5pt} 100 }    & MPSNR   & 10.48  & 17.65  & 22.51  & 27.81              & 26.82  & 27.86              & 28.02              & 27.41              & \textbf{29.48}     & \underline{28.04}  \\
		                                      & MSSIM   & 0.2888 & 0.5427 & 0.8141 & 0.9475             & 0.9322 & 0.9460             & \underline{0.9491} & 0.9375             & \textbf{0.9618}    & 0.9460             \\
		\hline
		\hline
		\multirow{2}{*}{\hspace{5pt} [0-15] } & MPSNR   & 33.32  & 34.62  & 37.22  & 39.89              & 40.04  & 37.40              & 40.77              & \textbf{43.72}     & \underline{41.83}  & 38.16              \\
		                                      & MSSIM   & 0.9551 & 0.9746 & 0.9903 & 0.9950             & 0.9951 & 0.9926             & 0.9958             & \textbf{0.9971}    & \underline{0.9968} & 0.9917             \\
		\hline
		\multirow{2}{*}{\hspace{5pt}[0-55] }  & MPSNR   & 22.45  & 26.11  & 29.04  & 38.37              & 33.36  & 32.55              & 34.31              & \underline{38.44}  & \textbf{39.28}     & 31.93              \\
		                                      & MSSIM   & 0.7450 & 0.8683 & 0.9504 & \underline{0.9934} & 0.9811 & 0.9780             & 0.9859             & 0.9925             & \textbf{0.9945}    & 0.9781             \\
		\hline
		\multirow{2}{*}{\hspace{5pt}[0-95] }  & MPSNR   & 18.18  & 23.06  & 25.77  & \underline{36.98}  & 30.07  & 29.21              & 30.80              & 35.84              & \textbf{37.20}     & 27.79              \\
		                                      & MSSIM   & 0.5889 & 0.7688 & 0.9033 & \underline{0.9914} & 0.9643 & 0.9589             & 0.9718             & 0.9877             & \textbf{0.9920}    & 0.9561             \\
		\hline
		\hline
		\multirow{2}{*}{\hspace{5pt}Corr. }   & MPSNR   & 28.48  & 30.50  & 33.69  & 37.96              & 37.77  & 36.56              & 38.54              & \underline{39.84}  & \textbf{40.79}     & 39.61              \\
		                                      & MSSIM   & 0.9085 & 0.9515 & 0.9637 & 0.9928             & 0.9921 & 0.9911             & 0.9934             & \underline{0.9944} & \textbf{0.9960}    & \underline{0.9944} \\
		\hline
		\hline
		\multirow{2}{*}{\hspace{5pt}Strip. }  & MPSNR   & 20.47  & 24.08  & 29.07  & 35.27              & 34.13  & 34.94              & 35.24              & \underline{35.25}  & \textbf{36.34}     & 34.50              \\
		                                      & MSSIM   & 0.7621 & 0.8672 & 0.9433 & 0.9877             & 0.9833 & \underline{0.9876} & \underline{0.9876} & 0.9874             & \textbf{0.9906}    & 0.9853             \\
		\bottomrule
	\end{tabular}
}

\end{table}
\begin{table}[H]
	\centering
	\captionof{table}{Denoising performances on Washington DC Mall with additional baselines.}\label{table:dcmallext}
	\resizebox{\textwidth}{!}{
	\begin{tabular}{c c c c c c c c c c c c c c c}
		\hline
		$\hspace{1pt}\sigma$ \hspace{1pt}  & Metrics & Noisy  & BM3D   & BM4D   & GLF    & LLRT   & NGMeet & 3D-ADNet & HSID-CNN & HSI-SDeCNN & SMDS-Net & QRNN3D & T3SC            \\ [0.5ex]
		\hline\hline
		\multirow{2}{*}{\hspace{5pt} 5  }  & MPSNR   & 34.31  & 35.10  & 41.13  & 39.57  & 41.83  & 37.57  & 42.08    & 41.68    & 39.98      & 42.83    & 43.42  & \textbf{43.85}  \\
		                                   & MSSIM   & 0.9821 & 0.9875 & 0.9962 & 0.9953 & 0.9968 & 0.9928 & 0.9968   & 0.9966   & 0.9954     & 0.9971   & 0.9973 & \textbf{0.9978} \\
		\hline
		\multirow{2}{*}{\hspace{5pt} 25 }  & MPSNR   & 20.70  & 24.51  & 31.08  & 35.25  & 34.95  & 35.38  & 33.78    & 33.05    & 33.44      & 35.64    & 35.04  & \textbf{36.74}  \\
		                                   & MSSIM   & 0.7688 & 0.8859 & 0.9690 & 0.9883 & 0.9863 & 0.9886 & 0.9825   & 0.9813   & 0.9822     & 0.9889   & 0.9864 & \textbf{0.9912} \\
		\hline
		\multirow{2}{*}{\hspace{5pt} 50 }  & MPSNR   & 15.25  & 20.80  & 26.69  & 31.77  & 30.94  & 31.88  & 29.73    & 28.96    & 29.61      & 31.76    & 31.72  & \textbf{33.12}  \\
		                                   & MSSIM   & 0.5314 & 0.7508 & 0.9220 & 0.9761 & 0.9704 & 0.9759 & 0.9587   & 0.9536   & 0.9608     & 0.9765   & 0.9741 & \textbf{0.9819} \\
		\hline
		\multirow{2}{*}{\hspace{5pt} 100 } & MPSNR   & 10.48  & 17.65  & 22.51  & 27.81  & 26.82  & 27.86  & 24.74    & 25.29    & 25.75      & 28.02    & 27.41  & \textbf{29.48}  \\
		                                   & MSSIM   & 0.2888 & 0.5427 & 0.8141 & 0.9475 & 0.9322 & 0.9460 & 0.9064   & 0.9014   & 0.9121     & 0.9491   & 0.9375 & \textbf{0.9618} \\
		\hline
	\end{tabular}
}

\end{table}

\paragraph{Study of statistical significance for the ICVL dataset.} In order to evaluate the statistical significance of our results, we
present some results in Table~\ref{table:std} for some of our models and baselines, by running models with five different random seeds.
Note that we did not conduct such a study for all results in this paper in order to keep the computational cost of the project reasonable.
The conclusions of the paper remain unchanged.

\begin{table}[H]
	\centering
	\captionof{table}{Denoising performances on ICVL with multiple seeds}\label{table:std}
	\def\arraystretch{1.2}
\resizebox{\textwidth}{!}{
	\begin{tabular}{c c c c c c c c c}
		\toprule
		$\hspace{1pt}\sigma$ \hspace{1pt}            & Metrics & Noisy                    & GLF                       & NGMeet                            & SMDS                     & QRNN3D                               & T3SC                                 & T3SC-SSL                             \\ [0.5ex]
		\hline\hline

		\multirow{2}{*}{\hspace{5pt}  5            } & MPSNR   & \mc{$34.47 \pm 0.01$}    & \mc{$51.25 \pm 0.01$}     & \mc{$\mathbf{52.74 \pm 0.01}$}    & \mc{$50.78 \pm 0.09$}    & \mc{$49.54 \pm 1.28$}                & \mc{$\underline{52.62 \pm0.01}$}     & \mc{$51.37 \pm 0.03$}                \\
		                                             & MSSIM   & \mc{$0.7619 \pm 0.0001$} & \mc{$0.9951 \pm 0.0001$}  & \mc{$\mathbf{0.9961 \pm 0.0001}$} & \mc{$0.9943 \pm 0.0001$} & \mc{$0.9924 \pm 0.0021$}             & \mc{$\underline{0.9960 \pm 0.0001}$} & \mc{$0.9952 \pm 0.0001$}             \\
		\hline
		\multirow{2}{*}{\hspace{5pt} 25 }            & MPSNR   & \mc{$21.43 \pm 0.01$}    & \mc{$43.16 \pm 0.01$}     & \mc{$\underline{44.74 \pm 0.01}$} & \mc{$42.63 \pm 0.11$}    & \mc{$44.20 \pm 0.16$}                & \mc{$\mathbf{45.37 \pm 0.02}$}       & \mc{$44.70 \pm 0.02$}                \\
		                                             & MSSIM   & \mc{$0.1548 \pm 0.0002$} & \mc{$0.9696 \pm 0.0001$}  & \mc{$0.9797 \pm 0.0001$}          & \mc{$0.9687 \pm 0.0009$} & \mc{$0.9780 \pm 0.0009$}             & \mc{$\mathbf{0.9825 \pm 0.0001}$}    & \mc{$\underline{0.9805 \pm 0.0001}$} \\
		\hline
		\multirow{2}{*}{\hspace{5pt} 50 }            & MPSNR   & \mc{$16.03 \pm 0.01$}    & \mc{$39.26 \pm 0.01$}     & \mc{$41.09 \pm 0.01$}             & \mc{$39.09 \pm 0.08$}    & \mc{$41.47 \pm 0.14$}                & \mc{$\mathbf{42.16 \pm 0.01}$}       & \mc{$\underline{41.62 \pm 0.01}$}    \\
		                                             & MSSIM   & \mc{$0.0503 \pm 0.0001$} & \mc{$0.9198 \pm  0.0002$} & \mc{$0.9603 \pm 0.0001$}          & \mc{$0.9359 \pm 0.0012$} & \mc{$0.9639 \pm 0.0012$}             & \mc{$\mathbf{0.9677 \pm 0.0001}$}    & \mc{$\underline{0.9648 \pm 0.0001}$} \\
		\hline
		\multirow{2}{*}{\hspace{5pt} 100 }           & MPSNR   & \mc{$10.85 \pm 0.01$}    & \mc{$34.78 \pm 0.01$}     & \mc{$37.55 \pm 0.01$}             & \mc{$35.59 \pm 0.04$}    & \mc{$38.38 \pm 0.60$}                & \mc{$\mathbf{38.99 \pm 0.01}$}       & \mc{$\underline{38.51 \pm 0.01}$}    \\
		                                             & MSSIM   & \mc{$0.0144 \pm 0.0001$} & \mc{$0.7981 \pm 0.0004$}  & \mc{$0.9312 \pm 0.0001$}          & \mc{$0.8781 \pm 0.0017$} & \mc{$\underline{0.9370 \pm 0.0114}$} & \mc{$\mathbf{0.9439 \pm 0.0002}$}    & \mc{$\underline{0.9397 \pm 0.0001}$} \\
		\hline
		\multirow{2}{*}{\hspace{5pt} [0-15] }        & MPSNR   & \mc{$33.94 \pm 0.09$}    & \mc{$50.68 \pm 0.11$}     & \mc{$41.57 \pm 0.14$}             & \mc{$48.00 \pm 0.13$}    & \mc{$\underline{52.10 \pm 0.12}$}    & \mc{$\mathbf{53.10 \pm 0.12}$}       & \mc{$51.21 \pm 0.11$}                \\
		                                             & MSSIM   & \mc{$0.6381 \pm 0.0013$} & \mc{$0.9950 \pm 0.0001$}  & \mc{$0.9065 \pm 0.0022$}          & \mc{$0.9899 \pm 0.0001$} & \mc{$\underline{0.9958 \pm 0.0001}$} & \mc{$\mathbf{0.9966 \pm 0.0001}$}    & \mc{$0.9955 \pm 0.0002$}             \\
		\hline
		\multirow{2}{*}{\hspace{5pt}[0-55] }         & MPSNR   & \mc{$23.41 \pm 0.09$}    & \mc{$44.41 \pm 0.12$}     & \mc{$32.93 \pm 0.09$}             & \mc{$41.42 \pm 0.18$}    & \mc{$\underline{47.26 \pm 0.12}$}    & \mc{$\mathbf{48.57 \pm 0.28}$}       & \mc{$46.47 \pm 0.23$}                \\
		                                             & MSSIM   & \mc{$0.2621 \pm 0.0025$} & \mc{$0.9820 \pm 0.0004$}  & \mc{$0.7534 \pm 0.0031$}          & \mc{$0.9593 \pm 0.0015$} & \mc{$\underline{0.9889 \pm 0.0004}$} & \mc{$\mathbf{0.9915 \pm 0.0005}$}    & \mc{$0.9856 \pm 0.0024$}             \\
		\hline
		\multirow{2}{*}{\hspace{5pt}[0-95] }         & MPSNR   & \mc{$19.11 \pm 0.09$}    & \mc{$41.62 \pm 0.11$}     & \mc{$29.40 \pm 0.12$}             & \mc{$38.86 \pm 0.06$}    & \mc{$\underline{44.07 \pm 0.08}$}    & \mc{$\mathbf{46.24 \pm 0.24}$}       & \mc{$43.98 \pm 0.46$}                \\
		                                             & MSSIM   & \mc{$0.1644 \pm 0.0031$} & \mc{$0.9667 \pm 0.0007$}  & \mc{$0.6601 \pm 0.0051$}          & \mc{$0.9352 \pm 0.0004$} & \mc{$\underline{0.9758 \pm 0.0003}$} & \mc{$\mathbf{0.9863 \pm 0.0005}$}    & \mc{$0.9735 \pm 0.0049$}             \\
		\bottomrule
	\end{tabular}
}

\end{table}

\paragraph{CAVE dataset.}
We report denoising performances of T3SC on the CAVE Dataset in Table ~\ref{table:cave}
To evaluate T3SC, the dataset was divided in four splits : three were used for training and one for testing.
The values reported for T3SC are averaged across all rotations of the test split.
\begin{table}[H]
	\centering
	\captionof{table}{Denoising performances on CAVE dataset with Gaussian noise.}\label{table:cave}
	\def\arraystretch{1.2}
\resizebox{0.5\textwidth}{!}{
	\begin{tabular}{c c c c c}
		\hline
		$\hspace{1pt}\sigma$ \hspace{1pt} & Metrics & Noisy & NGMeet & T3SC           \\ [0.5ex]
		\hline\hline
		5                                 & MPSNR   & 35.05 & 47.96  & \textbf{49.16} \\
		\hline
		25                                & MPSNR   & 21.99 & 42.44  & \textbf{42.77} \\
		\hline
		50                                & MPSNR   & 16.37 & 38.89  & \textbf{39.7}  \\
		\hline
		100                               & MPSNR   & 10.96 & 34.99  & \textbf{36.48} \\
		\hline
	\end{tabular}
}

\end{table}

\paragraph{Joint training across heterogeneous datasets.}
In Table~\ref{table:joint}, we study the problem of training a single model on
three different datasets, APEX, DC Mall, and Pavia, involving a different
number of channels. As mentioned in the paper, this model involves a common
second layer and a spectral dictionary per dataset. These result show that most
of the model parameters (which are present in the second layer) can in fact be shared
across datasets without significant loss of accuracy when compared to the training of
three different models (thus involving three times more parameters).

\begin{table}[H]
	\centering
	\captionof{table}{Results for joint training experiment}
	
\begin{tabular}{c |c | c c c c}
	\hline
	Training procedure                      & Model                   & Metrics    & APEX                 & DC Mall              & Pavia Center         \\ [0.5ex]
	\hline\hline
	\multirow{4}{*}{Independant trainings } & \multirow{2}{*}{QRNN3D} & \mc{MPSNR} & \mc{33.19}           & \mc{31.72}           & \mc{30.56}           \\
	                                        &                         & \mc{MSSIM} & \mc{0.9619}          & \mc{0.9741}          & \mc{0.9569}          \\
	                                        & \multirow{2}{*}{T3SC}   & \mc{MPSNR} & \mc{\textbf{34.91}}  & \mc{\textbf{33.12}}  & \mc{\textbf{31.32}}  \\
	                                        &                         & \mc{MSSIM} & \mc{\textbf{0.9730}} & \mc{\textbf{0.9819}} & \mc{\textbf{0.9617}} \\
	\hline
	\multirow{4}{*}{Joint training }        & \multirow{2}{*}{QRNN3D} & \mc{MPSNR} & \mc{31.95}           & \mc{30.97}           & \mc{29.12}           \\
	                                        &                         & \mc{MSSIM} & \mc{0.9501}          & \mc{0.9690}          & \mc{0.9428}          \\
	                                        & \multirow{2}{*}{T3SC}   & \mc{MPSNR} & \mc{\textbf{34.74}}  & \mc{\textbf{33.08}}  & \mc{\textbf{31.30}}  \\
	                                        &                         & \mc{MSSIM} & \mc{\textbf{0.9711}} & \mc{\textbf{0.9819}} & \mc{\textbf{0.9616}} \\
	\hline
\end{tabular}

	\label{table:joint}
\end{table}

\paragraph{Additional metrics.}
Additional metrics are provided for the ICVL and DCMall datasets, respectively in Tables~\ref{table:metric1} and~\ref{table:metric2}.
The conclusions of the paper are unchanged.
\begin{table}[H]
	\centering
	\def\arraystretch{1.2}
\resizebox{\textwidth}{!}{
	\begin{tabular}{c c c c c c c c c c c c}
		\hline
		$\hspace{1pt}\sigma$ \hspace{1pt}     & Metrics     & Noisy       & BM3D        & BM4D        & GLF         & LLRT                    & NGMeet                  & SMDS        & QRNN3D                  & T3SC                    & T3SC-SSL                \\ [0.5ex]
		\hline\hline
		\multirow{3}{*}{\hspace{5pt}  5}      & \mc{MFSIM}  & \mc{0.9953} & \mc{0.9978} & \mc{0.9986} & \mc{0.9994} & \mc{\underline{0.9995}} & \mc{\textbf{0.9996}}    & \mc{0.9993} & \mc{0.9987}             & \mc{\textbf{0.9996}}    & \mc{\underline{0.9995}} \\
		                                      & \mc{MERGAS} & \mc{6.18}   & \mc{1.48}   & \mc{1.10}   & \mc{0.84}   & \mc{0.7740}             & \mc{\textbf{0.69}}      & \mc{0.87}   & \mc{1.14}               & \mc{\underline{0.70}}   & \mc{0.83}               \\
		                                      & \mc{MSAM}   & \mc{0.2460} & \mc{0.0518} & \mc{0.0390} & \mc{0.0267} & \mc{0.0229}             & \mc{\textbf{0.0211}}    & \mc{0.0307} & \mc{0.0412}             & \mc{\underline{0.0223}} & \mc{0.0286}             \\
		\hline
		\multirow{3}{*}{\hspace{5pt} 25 }     & \mc{MFSIM}  & \mc{0.9218} & \mc{0.9773} & \mc{0.9829} & \mc{0.9944} & \mc{0.9942}             & \mc{0.9954}             & \mc{0.9921} & \mc{\underline{0.9967}} & \mc{\textbf{0.9970}}    & \mc{0.9965}             \\
		                                      & \mc{MERGAS} & \mc{27.33}  & \mc{3.86}   & \mc{3.21}   & \mc{2.13}   & \mc{2.19}               & \mc{\underline{1.77}}   & \mc{2.20}   & \mc{1.86}               & \mc{\textbf{1.65}}      & \mc{1.79}               \\
		                                      & \mc{MSAM}   & \mc{0.5989} & \mc{0.1286} & \mc{0.1005} & \mc{0.0595} & \mc{0.0459}             & \mc{\textbf{0.0384}}    & \mc{0.0717} & \mc{0.0537}             & \mc{\underline{0.0406}} & \mc{0.0501}             \\
		\hline
		\multirow{3}{*}{\hspace{5pt} 50 }     & \mc{MFSIM}  & \mc{0.8100} & \mc{0.9488} & \mc{0.9488} & \mc{0.9851} & \mc{0.9851}             & \mc{0.9863}             & \mc{0.9782} & \mc{\textbf{0.9928}}    & \mc{\underline{0.9925}} & \mc{0.9914}             \\
		                                      & \mc{MERGAS} & \mc{51.48}  & \mc{5.88}   & \mc{5.88}   & \mc{3.33}   & \mc{3.92}               & \mc{2.71}               & \mc{3.33}   & \mc{\underline{2.50}}   & \mc{\textbf{2.40}}      & \mc{2.56}               \\
		                                      & \mc{MSAM}   & \mc{0.7546} & \mc{0.1964} & \mc{0.1964} & \mc{0.1029} & \mc{0.0682}             & \mc{\textbf{0.0505}}    & \mc{0.1033} & \mc{0.0571}             & \mc{\underline{0.0549}} & \mc{0.0663}             \\
		\hline
		\multirow{3}{*}{\hspace{5pt} 100 }    & \mc{MFSIM}  & \mc{0.6471} & \mc{0.8942} & \mc{0.9008} & \mc{0.9679} & \mc{0.9637}             & \mc{0.9661}             & \mc{0.9456} & \mc{\textbf{0.9835}}    & \mc{\underline{0.9824}} & \mc{0.9805}             \\
		                                      & \mc{MERGAS} & \mc{95.97}  & \mc{9.11}   & \mc{7.96}   & \mc{5.59}   & \mc{6.22}               & \mc{4.08}               & \mc{5.04}   & \mc{\underline{4.20}}   & \mc{\textbf{3.46}}      & \mc{3.66}               \\
		                                      & \mc{MSAM}   & \mc{0.8619} & \mc{0.2984} & \mc{0.2228} & \mc{0.1847} & \mc{0.0919}             & \mc{\textbf{0.0679}}    & \mc{0.1441} & \mc{0.1009}             & \mc{\underline{0.0761}} & \mc{0.0889}             \\
		\hline\hline
		\multirow{3}{*}{\hspace{5pt} [0-15] } & \mc{MFSIM}  & \mc{0.9876} & \mc{0.9954} & \mc{0.9963} & \mc{0.9991} & \mc{0.9985}             & \mc{0.9965}             & \mc{0.9984} & \mc{\underline{0.9995}} & \mc{\textbf{0.9996}}    & \mc{0.9993}             \\
		                                      & \mc{MERGAS} & \mc{10.11}  & \mc{1.91}   & \mc{2.07}   & \mc{0.98}   & \mc{1.17}               & \mc{4.53}               & \mc{1.20}   & \mc{\underline{0.79}}   & \mc{\textbf{0.69}}      & \mc{0.91}               \\
		                                      & \mc{MSAM}   & \mc{0.3412} & \mc{0.0680} & \mc{0.0672} & \mc{0.0328} & \mc{0.0311}             & \mc{0.1772}             & \mc{0.0408} & \mc{\underline{0.0265}} & \mc{\textbf{0.0234}}    & \mc{0.0322}             \\
		\hline
		\multirow{3}{*}{\hspace{5pt}[0-55] }  & \mc{MFSIM}  & \mc{0.9087} & \mc{0.9743} & \mc{0.9768} & \mc{0.9950} & \mc{0.9900}             & \mc{0.9755}             & \mc{0.9890} & \mc{\underline{0.9984}} & \mc{\textbf{0.9985}}    & \mc{0.9978}             \\
		                                      & \mc{MERGAS} & \mc{33.34}  & \mc{4.17}   & \mc{4.73}   & \mc{2.07}   & \mc{3.02}               & \mc{14.69}              & \mc{2.50}   & \mc{\underline{1.39}}   & \mc{\textbf{1.20}}      & \mc{1.52}               \\
		                                      & \mc{MSAM}   & \mc{0.6478} & \mc{0.1443} & \mc{0.1412} & \mc{0.0687} & \mc{0.0636}             & \mc{\underline{0.4086}} & \mc{0.0784} & \mc{0.0427}             & \mc{\textbf{0.0370}}    & \mc{0.0502}             \\
		\hline
		\multirow{3}{*}{\hspace{5pt}[0-95] }  & \mc{MFSIM}  & \mc{0.8291} & \mc{0.9524} & \mc{0.9560} & \mc{0.9911} & \mc{0.9798}             & \mc{0.9536}             & \mc{0.9772} & \mc{\underline{0.9969}} & \mc{\textbf{0.9972}}    & \mc{0.9962}             \\
		                                      & \mc{MERGAS} & \mc{54.92}  & \mc{5.83}   & \mc{6.73}   & \mc{2.86}   & \mc{4.64}               & \mc{24.82}              & \mc{3.46}   & \mc{2.17}               & \mc{\textbf{1.58}}      & \mc{\underline{1.92}}   \\
		                                      & \mc{MSAM}   & \mc{0.7720} & \mc{0.2001} & \mc{0.1928} & \mc{0.0992} & \mc{0.0813}             & \mc{0.5574}             & \mc{0.1042} & \mc{0.0622}             & \mc{\textbf{0.0471}}    & \mc{\underline{0.0615}} \\
		\hline\hline
		\multirow{3}{*}{\hspace{5pt}Corr. }   & \mc{MFSIM}  & \mc{0.9704} & \mc{0.9902} & \mc{0.9923} & \mc{0.9981} & \mc{0.9968}             & \mc{0.9919}             & \mc{0.9969} & \mc{\underline{0.9990}} & \mc{\textbf{0.9991}}    & \mc{0.9988}             \\
		                                      & \mc{MERGAS} & \mc{14.20}  & \mc{2.61}   & \mc{3.74}   & \mc{1.46}   & \mc{1.63}               & \mc{6.37}               & \mc{1.55}   & \mc{\underline{1.12}}   & \mc{\textbf{1.02}}      & \mc{1.15}               \\
		                                      & \mc{MSAM}   & \mc{0.4617} & \mc{0.0934} & \mc{0.1540} & \mc{0.0468} & \mc{0.0416}             & \mc{0.2550}             & \mc{0.0515} & \mc{\underline{0.0316}} & \mc{\textbf{0.0291}}    & \mc{0.0367}             \\
		\hline\hline
		\multirow{3}{*}{\hspace{5pt}Strip. }  & \mc{MFSIM}  & \mc{0.9068} & \mc{0.9579} & \mc{0.9736} & \mc{0.9926} & \mc{0.9871}             & \mc{0.9880}             & \mc{0.9900} & \mc{\textbf{0.9968}}    & \mc{\underline{0.9965}} & \mc{0.9956}             \\
		                                      & \mc{MERGAS} & \mc{28.14}  & \mc{7.65}   & \mc{4.65}   & \mc{2.52}   & \mc{4.34}               & \mc{4.32}               & \mc{2.44}   & \mc{\underline{1.78}}   & \mc{\textbf{1.77}}      & \mc{2.00}               \\
		                                      & \mc{MSAM}   & \mc{0.6067} & \mc{0.2197} & \mc{0.1442} & \mc{0.0764} & \mc{0.1272}             & \mc{0.1298}             & \mc{0.0790} & \mc{\textbf{0.0439}}    & \mc{\underline{0.0534}} & \mc{0.0631}             \\
		\hline
	\end{tabular}
}

	\captionof{table}{Additional metrics on ICVL}\label{table:metric1}
\end{table}

\begin{table}[H]
	\centering
	\resizebox{\textwidth}{!}{
	\begin{tabular}{c c c c c c c c c c c c c}
		\hline
		$\hspace{1pt}\sigma$ \hspace{1pt}     & Metrics     & Noisy       & BM3D        & BM4D        & GLF                     & LLRT        & NGMeet                  & SMDS                  & QRNN3D                  & T3SC                    & T3SC-SSL           \\ [0.5ex]
		\hline\hline
		\multirow{4}{*}{\hspace{5pt} 5  }     & \mc{MFSIM}  & \mc{0.9534} & \mc{0.9578} & \mc{0.9772} & \mc{0.9824}             & \mc{0.9817} & \mc{0.9785}             & \mc{0.9802}           & \mc{\textbf{0.9824}}    & \mc{\underline{0.9814}} & 0.9804             \\
		                                      & \mc{MERGAS} & \mc{3.12}   & \mc{2.84}   & \mc{1.50}   & \mc{1.96}               & \mc{1.46}   & \mc{2.50}               & \mc{1.38}             & \mc{\underline{1.26}}   & \mc{\textbf{1.19}}      & 1.54               \\
		                                      & \mc{MSAM}   & \mc{0.0862} & \mc{0.0775} & \mc{0.0427} & \mc{0.0495}             & \mc{0.0395} & \mc{0.0569}             & \mc{0.0373}           & \mc{\underline{0.0349}} & \mc{\textbf{0.0329}}    & 0.0425             \\
		\hline
		\multirow{4}{*}{\hspace{5pt} 25 }     & \mc{MFSIM}  & \mc{0.8213} & \mc{0.8676} & \mc{0.9394} & \mc{\underline{0.9661}} & \mc{0.9629} & \mc{0.9655}             & \mc{0.9639}           & \mc{0.9614}             & \mc{\textbf{0.9673}}    & 0.9648             \\
		                                      & \mc{MERGAS} & \mc{14.96}  & \mc{9.50}   & \mc{4.55}   & \mc{2.91}               & \mc{3.31}   & \mc{2.94}               & \mc{2.87}             & \mc{3.08}               & \mc{\textbf{2.50}}      & \underline{2.77}   \\
		                                      & \mc{MSAM}   & \mc{0.3087} & \mc{0.1753} & \mc{0.1044} & \mc{0.0684}             & \mc{0.0726} & \mc{0.0671}             & \mc{0.0676}           & \mc{0.0709}             & \mc{\textbf{0.0599}}    & \underline{0.0668} \\
		\hline
		\multirow{4}{*}{\hspace{5pt} 50 }     & \mc{MFSIM}  & \mc{0.7174} & \mc{0.7861} & \mc{0.8974} & \mc{\underline{0.9495}} & \mc{0.9439} & \mc{0.9484}             & \mc{0.9464}           & \mc{0.9487}             & \mc{\textbf{0.9542}}    & 0.9465             \\
		                                      & \mc{MERGAS} & \mc{28.00}  & \mc{14.51}  & \mc{7.44}   & \mc{4.24}               & \mc{4.89}   & \mc{4.28}               & \mc{4.45}             & \mc{4.38}               & \mc{\textbf{3.68}}      & \underline{4.23}   \\
		                                      & \mc{MSAM}   & \mc{0.4785} & \mc{0.2175} & \mc{0.1438} & \mc{0.0890}             & \mc{0.0925} & \mc{\underline{0.0864}} & \mc{0.0944}           & \mc{0.0880}             & \mc{\textbf{0.0768}}    & 0.0934             \\
		\hline
		\multirow{4}{*}{\hspace{5pt} 100 }    & \mc{MFSIM}  & \mc{0.6000} & \mc{0.6821} & \mc{0.8240} & \mc{0.9188}             & \mc{0.9065} & \mc{\underline{0.9209}} & \mc{0.9170}           & \mc{0.9100}             & \mc{\textbf{0.9329}}    & 0.9163             \\
		                                      & \mc{MERGAS} & \mc{48.42}  & \mc{20.83}  & \mc{11.98}  & \mc{6.54}               & \mc{7.58}   & \mc{6.66}               & \mc{\underline{6.52}} & \mc{7.01}               & \mc{\textbf{5.51}}      & \underline{6.52}   \\
		                                      & \mc{MSAM}   & \mc{0.6566} & \mc{0.2700} & \mc{0.1939} & \mc{0.1183}             & \mc{0.1193} & \mc{\underline{0.1147}} & \mc{0.1205}           & \mc{0.1297}             & \mc{\textbf{0.0977}}    & 0.1265             \\
		\hline\hline
		\multirow{4}{*}{\hspace{5pt} [0-15] } & \mc{MFSIM}  & \mc{0.9338} & \mc{0.9455} & \mc{0.9690} & \mc{\textbf{0.9831}}    & \mc{0.9774} & \mc{0.9761}             & \mc{0.9787}           & \mc{\underline{0.9828}} & \mc{0.9782}             & 0.9679             \\
		                                      & \mc{MERGAS} & \mc{5.42}   & \mc{4.29}   & \mc{2.29}   & \mc{2.10}               & \mc{1.89}   & \mc{2.53}               & \mc{1.68}             & \mc{\textbf{1.36}}      & \mc{\underline{1.48}}   & 2.34               \\
		                                      & \mc{MSAM}   & \mc{0.1358} & \mc{0.1052} & \mc{0.0610} & \mc{0.0509}             & \mc{0.0487} & \mc{0.0582}             & \mc{0.0438}           & \mc{\textbf{0.0368}}    & \mc{\underline{0.0395}} & 0.0624             \\
		\hline
		\multirow{4}{*}{\hspace{5pt}[0-55] }  & \mc{MFSIM}  & \mc{0.8196} & \mc{0.8642} & \mc{0.9261} & \mc{\textbf{0.9766}}    & \mc{0.9554} & \mc{0.9523}             & \mc{0.9603}           & \mc{0.9714}             & \mc{\underline{0.9748}} & 0.9507             \\
		                                      & \mc{MERGAS} & \mc{18.46}  & \mc{10.41}  & \mc{5.56}   & \mc{\underline{2.37}}   & \mc{3.86}   & \mc{4.19}               & \mc{3.22}             & \mc{\underline{2.37}}   & \mc{\textbf{2.05}}      & 4.73               \\
		                                      & \mc{MSAM}   & \mc{0.3563} & \mc{0.1879} & \mc{0.1171} & \mc{\underline{0.0572}} & \mc{0.0798} & \mc{0.0961}             & \mc{0.0731}           & \mc{0.0581}             & \mc{\textbf{0.0518}}    & 0.1226             \\
		\hline
		\multirow{4}{*}{\hspace{5pt}[0-95] }  & \mc{MFSIM}  & \mc{0.7471} & \mc{0.8057} & \mc{0.8837} & \mc{\textbf{0.9725}}    & \mc{0.9377} & \mc{0.9339}             & \mc{0.9473}           & \mc{0.9613}             & \mc{\underline{0.9689}} & 0.9370             \\
		                                      & \mc{MERGAS} & \mc{29.42}  & \mc{14.25}  & \mc{8.15}   & \mc{\underline{2.68}}   & \mc{5.36}   & \mc{6.14}               & \mc{4.60}             & \mc{3.07}               & \mc{\textbf{2.50}}      & 6.95               \\
		                                      & \mc{MSAM}   & \mc{0.4899} & \mc{0.2274} & \mc{0.1466} & \mc{\underline{0.0632}} & \mc{0.0973} & \mc{0.1262}             & \mc{0.0962}           & \mc{0.0719}             & \mc{\textbf{0.0604}}    & 0.1520             \\
		\hline\hline
		\multirow{4}{*}{\hspace{5pt} Corr. }  & \mc{MFSIM}  & \mc{0.9028} & \mc{0.9229} & \mc{0.9519} & \mc{\underline{0.9783}} & \mc{0.9713} & \mc{0.9693}             & \mc{0.9721}           & \mc{\textbf{0.9790}}    & \mc{0.9768}             & 0.9744             \\
		                                      & \mc{MERGAS} & \mc{8.25}   & \mc{5.91}   & \mc{4.07}   & \mc{2.29}               & \mc{2.44}   & \mc{2.67}               & \mc{2.10}             & \mc{\underline{1.92}}   & \mc{\textbf{1.65}}      & 1.97               \\
		                                      & \mc{MSAM}   & \mc{0.2049} & \mc{0.1368} & \mc{0.1106} & \mc{0.0559}             & \mc{0.0593} & \mc{0.0661}             & \mc{0.0540}           & \mc{\underline{0.0481}} & \mc{\textbf{0.0436}}    & 0.0527             \\
		\hline\hline
		\multirow{4}{*}{\hspace{5pt} Strip. } & \mc{MFSIM}  & \mc{0.8177} & \mc{0.8621} & \mc{0.9365} & \mc{\textbf{0.9663}}    & \mc{0.9604} & \mc{0.9649}             & \mc{0.9639}           & \mc{0.9619}             & \mc{\underline{0.9651}} & 0.9582             \\
		                                      & \mc{MERGAS} & \mc{15.38}  & \mc{10.20}  & \mc{4.84}   & \mc{3.00}               & \mc{3.55}   & \mc{3.09}               & \mc{\underline{2.99}} & \mc{3.02}               & \mc{\textbf{2.62}}      & 3.26               \\
		                                      & \mc{MSAM}   & \mc{0.3152} & \mc{0.1886} & \mc{0.1101} & \mc{\underline{0.0698}} & \mc{0.0794} & \mc{0.0705}             & \mc{0.0700}           & \mc{0.0702}             & \mc{\textbf{0.0623}}    & 0.0795             \\
		\hline
	\end{tabular}
}

	\captionof{table}{Additional metrics on DCMall}\label{table:metric2}
\end{table}

\paragraph{Ablation studies.} In this paragraph, we present different ablation studies, demonstrating in Table~\ref{table:multilayer} that our two-layer model outperforms single-layer models.
We also demonstrate the usefulness of our variant with weights~$\beta_j$ in Table~\ref{table:beta} when the noise variance varies a lot between different bands.
\begin{table}[H]
	\centering
	\begin{tabular}{c c c c c}
	\hline
	Metrics     & Noisy       & Spec        & SpecSpat    & Spec + SpecSpat      \\ [0.5ex]
	\hline\hline
	\mc{MPSNR}  & \mc{16.03}  & \mc{30.96}  & \mc{40.13}  & \mc{\textbf{42.17}}  \\
	\mc{MSSIM}  & \mc{0.0502} & \mc{0.6884} & \mc{0.9533} & \mc{\textbf{0.9677}} \\
	\mc{MFSIM}  & \mc{0.8100} & \mc{0.9708} & \mc{0.9849} & \mc{\textbf{0.9925}} \\
	\mc{MERGAS} & \mc{51.48}  & \mc{8.84}   & \mc{3.00}   & \mc{\textbf{2.39}}   \\
	\mc{MSAM}   & \mc{0.7546} & \mc{0.1300} & \mc{0.1021} & \mc{\textbf{0.0547}} \\
	\hline
\end{tabular}

	\captionof{table}{Combination of sparse coding layers: we denote by \textit{Spec} the Spectral Sparse Coding layer and by \textit{SpecSpat} the Spectral-Spatial Sparse Coding layer. This experiment was run on ICVL with $\sigma=50$.}
	\label{table:multilayer}
\end{table}

\begin{table}[H]
	\centering
	\captionof{table}{Our model without/with band-wise noise estimator (NE) on ICVL with band-dependent Gaussian noise and stripes noise}\label{table:beta}
	\resizebox{.5\textwidth}{!}{
	\begin{tabular}{c c c c }
		\hline
		                                      & Metrics    & T3SC        & T3SC + NE            \\
		\hline\hline
    \multirow{2}{*}{\hspace{5pt} [0-15] } & \mc{MPSNR} & \mc{52.85}  & \mc{\textbf{53.31}}  \\
                                          & \mc{MSSIM} & \mc{0.9963} & \mc{\textbf{0.9967}} \\
    \hline                                                  \hline
    \multirow{2}{*}{\hspace{5pt}[0-55] }  & \mc{MPSNR} & \mc{47.39}  & \mc{\textbf{48.64}}  \\
                                          & \mc{MSSIM} & \mc{0.9890} & \mc{\textbf{0.9911}} \\
    \hline
		\multirow{2}{*}{\hspace{5pt}[0-95] }  & \mc{MPSNR} & \mc{44.92}  & \mc{\textbf{46.30}}  \\
		                                      & \mc{MSSIM} & \mc{0.9821} & \mc{\textbf{0.9859}} \\
		\hline
		\multirow{2}{*}{\hspace{5pt}Strip. }  & \mc{MPSNR} & \mc{44.68}  & \mc{\textbf{44.74}}  \\
		                                      & \mc{MSSIM} & \mc{0.9801} & \mc{\textbf{0.9805}} \\
		\hline
	\end{tabular}
}

\end{table}

\section{Visual Examples}
\begin{figure}[H]
	\input{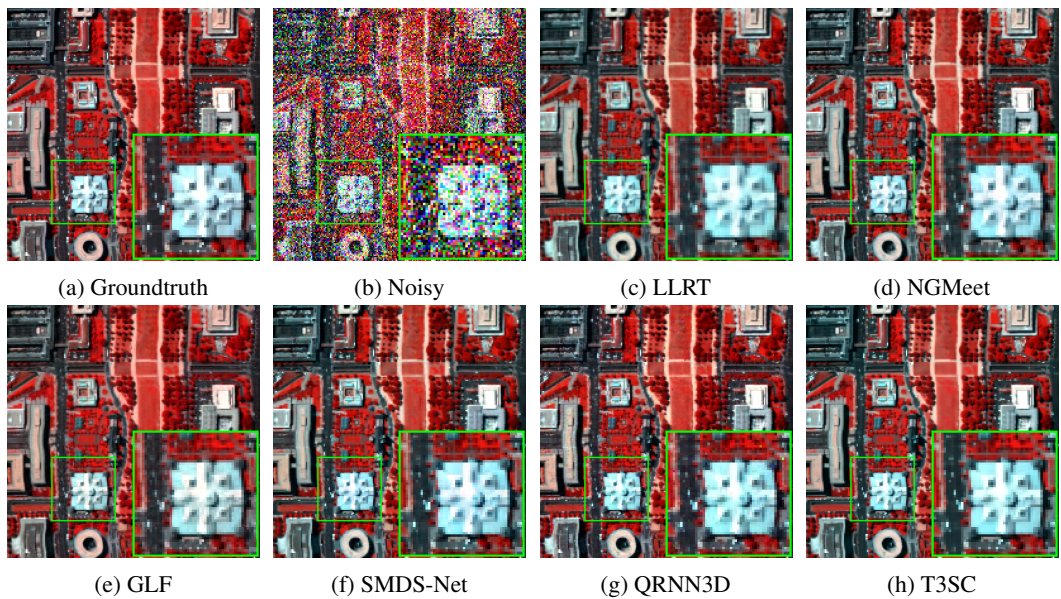}
	\caption{Simulated Gaussian noise ($\sigma=100$) on DCMall}\label{fig:1}
\end{figure}

\begin{figure}[H]
	\input{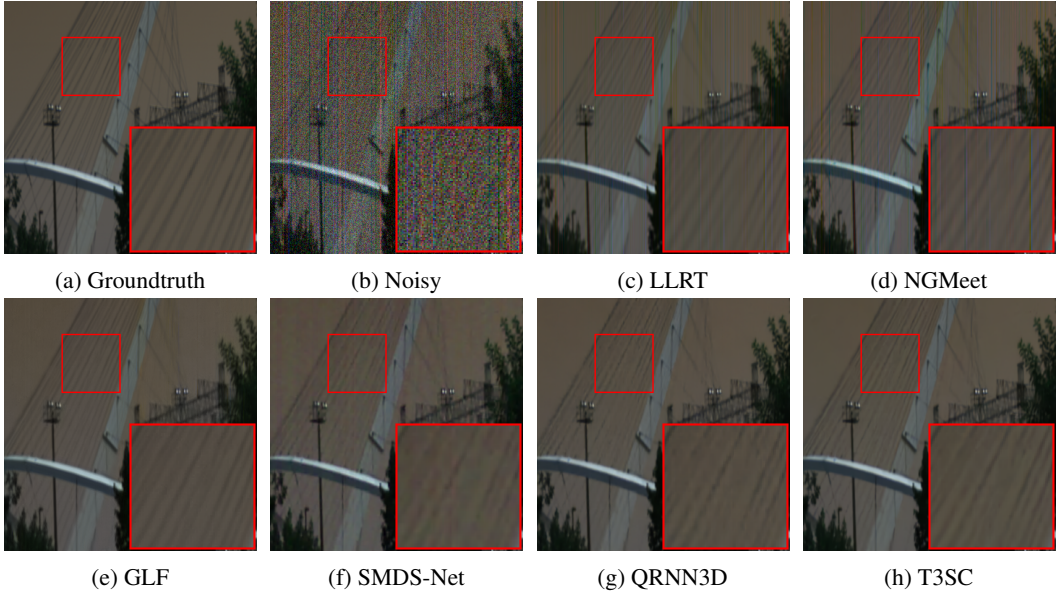}
	\caption{Visual results for the denoising experiment with stripes noise on ICVL with bands 9, 15, 28.}\label{fig:2}
\end{figure}

Finally, we show additional visual examples  in Figure~\ref{fig:1} and ~\ref{fig:2}.

\section{GPU ressources}
The total number of GPU hours involved in this project is around 19k hours on NVIDIA Tesla V100 16Go, including preliminary experiments, model design, final experiments and running baseline methods.

\end{document}